\documentclass[11pt,a4paper]{article}
\usepackage{amsmath,amssymb,graphicx,epsfig,cite,color,hyperref}
\usepackage[left=2.7cm,right=2.7cm,top=2.5cm,bottom=2.5cm]{geometry}
\usepackage{xcolor}

\def\smallfrac#1#2{\hbox{${{#1}\over {#2}}$}}
\newcommand{\be}{\begin{equation}}
\newcommand{\ee}{\end{equation}}
\def\bea#1\eea{\begin{align}#1\end{align}}
\newcommand{\bi}{\begin{itemize}}
\newcommand{\ei}{\end{itemize}}
\newcommand{\ben}{\begin{enumerate}}
\newcommand{\een}{\end{enumerate}}

\newcommand{\lp}{\left(}
\newcommand{\rp}{\right)}
\newcommand{\as}{\alpha_s}

\def\frac#1#2{{{#1}\over {#2}}}
\def\gsim{\mathrel{\rlap{\lower4pt\hbox{\hskip1pt$\sim$}}
    \raise1pt\hbox{$>$}}}         
\def\lsim{\mathrel{\rlap{\lower4pt\hbox{\hskip1pt$\sim$}}
    \raise1pt\hbox{$<$}}}         

\newcommand{\Li}{{\rm Li}}

\newcommand{\draft}[1]{}

\definecolor{grey}{rgb}{0.5,0.5,0.5}

\newcommand{\Ord}{\mathcal{O}}
\newcommand{\plus}[1]{\left[ #1 \right]_+}
\def\({\left(}
\def\){\right)}
\def\[{\left[}
\def\]{\right]}

\graphicspath{{plots/}}

\begin{document}

\vspace{-2.0cm}
\begin{flushright}
Edinburgh 2015/14\\
CERN-PH-TH/2015-176\\
TIF-UNIMI-2015-11\\
OUTP-15-20P \\
\end{flushright}

\begin{center}
{\Large \bf Intrinsic charm in a
  matched general-mass  scheme}
\vspace{.7cm}

Richard~D.~Ball$^{1}$, Valerio~Bertone$^{2,3}$,  Marco Bonvini$^{3}$,
Stefano~Forte$^4$,\\[0.1cm] Patrick Groth Merrild$^{1}$,  Juan~Rojo$^{3}$, and Luca Rottoli$^{3}$

\vspace{.3cm}
{\it ~$^1$ The Higgs Centre for Theoretical Physics, University of Edinburgh,\\
JCMB, KB, Mayfield Rd, Edinburgh EH9 3JZ, Scotland\\
~$^2$  PH Department, TH Unit, CERN, CH-1211 Geneva 23, Switzerland\\
~$^3$ Rudolf Peierls Centre for Theoretical Physics, 1 Keble Road,\\ University of Oxford, OX1 3NP Oxford, UK\\
~$^4$ TIF Lab, Dipartimento di Fisica, Universit\`a di Milano and\\
INFN, Sezione di Milano,\\ Via Celoria 16, I-20133 Milano, Italy\\
}
\end{center}

\vspace{0.1cm}

\begin{center}
  {\bf \large Abstract}
\end{center}
The FONLL general-mass variable-flavour number scheme provides a
framework for the matching of a calculation in which a heavy quark is
treated as a massless parton to one in which the mass dependence is
retained throughout. We describe how the usual formulation of FONLL
can be extended in such a way that the heavy quark parton distribution
functions are freely parameterized at some initial scale, rather than
being generated entirely perturbatively. We specifically consider the
case of deep-inelastic scattering, in view of applications to PDF
determination, and the possible impact of a fitted charm quark
distribution on $F_2^c$ is assessed.

\clearpage
In the perturbative computation of hard processes involving heavy quarks,
it is usually assumed that the
heavy quark content of colliding hadrons  is generated perturbatively,
namely, 
heavy quarks are generated from radiation by light partons.
This assumption  may
be unsatisfactory both for reasons of principle and of practice. As a
matter of principle, an ``intrinsic'' heavy quark
component~\cite{Brodsky:1980pb} may well be non-zero, especially in
the case of charm (see Ref.~\cite{Brodsky:2015fna} for a recent review).
Such intrinsic charm component might have observable consequences
at the LHC in processes like $\gamma+c$~\cite{Stavreva:2010mw,Bednyakov:2013zta} or
open charm production~\cite{Kniehl:2009ar}.
Also, in practice, if 
the heavy quark is assigned a parton distribution (PDF), as required
for accurate collider phenomenology~\cite{Thorne:2012az,Ball:2013gsa},
and this PDF is generated perturbatively, it will in general depend on
the choice of scale at which the perturbative boundary condition is
imposed. In a matched calculation this dependence will disappear at
high enough perturbative orders, but at low orders it might be
non-negligible in practice. 

Both problems
are solved by introducing a fitted heavy quark PDF, which can describe
a possible non-perturbative intrinsic component, and also, reabsorb in
the initial condition the dependence on the choice of starting scale
of the perturbative component.
It is the purpose
of the present paper to explain how the so-called FONLL approach of
Ref.~\cite{Forte:2010ta}, for the treatment of heavy
quarks with inclusion both of mass dependence, and resummation of
collinear logs,
 can be generalized to include such a
fitted heavy quark PDF.

The FONLL approach, originally proposed in Ref.~\cite{Cacciari:1998it},
specifically applied there to
heavy quark production in hadronic collisions, and generalized to
deep-inelastic scattering in Ref.~\cite{Forte:2010ta} (and more
recently to Higgs production in bottom quark
fusion~\cite{Forte:2015hba}), is a general-mass, variable-flavour number
(GM-VFN) scheme. Such schemes are designed to deal with the
fact that 
hard processes involving heavy quarks can be computed in perturbative QCD
using different renormalization and
factorization schemes: a massive, or decoupling
scheme, in which the heavy quark does not contribute to the running
of $\alpha_s$ or the DGLAP evolution equation, and it appears as a
massive field in the computation of hard cross-sections;
and a
massless scheme, in which the heavy quark is treated as a massless
parton.
In the former scheme, the mass dependence of the hard cross-section
is kept, but logs of the hard scale over the heavy quark mass are only
included to finite order, while in the latter scheme these logs are resummed
to all orders to some logarithmic accuracy through perturbative evolution, but heavy quark mass
effects are neglected.  For simplicity, we will henceforth refer to
the former as a three-flavour scheme (3FS), and to the latter as a four-flavour scheme (4FS),
which we will respectively take as synonyms of massive and massless scheme 
(though all we say would apply equally to four and five, or five and
six flavours). 

In GM-VFN schemes, the information contained in the
three- and four-flavour schemes are combined through a suitable
matching procedure. As the problem arises each time a 
heavy-quark threshold is crossed, the matching is
performed each time this happens.
The FONLL scheme has the dual
advantage that it can be generally
applied to any hard electro- or hadro-production process, and also, that
it allows for the combination of a three- and four-flavour 
computation each performed  at any desired perturbative order (fixed, in the
former case, and logarithmically resummed, in the latter).
Further GM-VFN schemes
include ACOT~\cite{Aivazis:1993pi,Collins:1998rz,Guzzi:2011ew} (and its variants
S-ACOT~\cite{Kramer:2000hn} and S-ACOT-$\chi$~\cite{Tung:2001mv})  and 
TR~\cite{Thorne:1997ga,Thorne:1997uu} (and its variant TR'~\cite{Thorne:2006qt,Thorne:2012az}), both of which have been mostly developed in the
context of deep-inelastic scattering (see Ref.~\cite{LHhq} for 
detailed comparisons), though applications of GM-VFN schemes to LHC
processes have been presented very recently~\cite{Han:2014nja,Bonvini:2015pxa}.

The possibility of including an intrinsic
charm component in a global PDF fit has been considered
previously by the CT
collaboration~\cite{Pumplin:2007wg,Dulat:2013hea} in the context of
the ACOT scheme. In these references,
however, the fitted PDF was only introduced in massless
contributions.
This may bias phenomenological conclusions, and, perhaps more
importantly, it does not allow for a fully consistent treatment of
the interplay between the fitted charm contribution to the fixed-order
and resummed computations.
A determination of intrinsic charm which instead only uses the
fixed-flavour number scheme  (i.e.\ a 3FS throughout)  has been
presented recently in Ref.~\cite{Jimenez-Delgado:2014zga}.

The basic idea of the FONLL method consists of expanding out the
massless-scheme computation in powers of the strong coupling
$\alpha_s$, and replacing a finite number of terms with their
massive-scheme counterparts. The result then has at the  massive level
the fixed-order accuracy which
corresponds to the number of massive orders which have been
included (``FO'', standing for fixed order), and  at the massless
level the same logarithmic accuracy as the starting 4FS
computation (``NLL'', standing for next$^k$-to-leading
logarithmic\footnote{The name 'FONLL' is perhaps a misnomer,
  as it suggests that the resummed calculation is necessarily NLL,
  while in actual fact it can be performed at any logarithmic
  order, but we stick to it for historical reasons.}).
 The only technical complication of the method is that the starting
three- and four-flavour scheme computations are performed in different
renormalization and factorization schemes. 
The difficulty is overcome by re-expressing both
$\alpha_s$ and the PDFs of the 3FS (which thus have
$n_f=3$ in the running of $\alpha_s$ and the evolution of PDFs) in
terms of those of the 4FS. This must be done order by
order in perturbation theory, to the desired accuracy of the computation.

The generic form of deep-inelastic structure functions in the FONLL
approach is
\begin{equation} \label{eq:fonll}
F (x, Q^2) =  F^{(3)} (x,Q^2) + F^{(4)} (x, Q^2) - F^{(3,0)} (x,Q^2)\, ,
\end{equation}
where the three- and four-flavour scheme structure functions are
respectively given by 
\begin{align}
 F^{(3)} (x, Q^2)& = x \int_x^1 \frac{dy}{y} \sum_{i = g, q, \bar{q}}
  C_i^{(3)} \left( \frac{x}{y}, \frac{Q^2}{m_h^2}, \alpha_s^{(3)} (Q^2) \right)
  f_i^{(3)} (y, Q^2)\nonumber\\&=\sum_{i = g, q, \bar{q}}
  C_i^{(3)} \left(\frac{Q^2}{m_h^2}, \alpha_s^{(3)} (Q^2) \right)\otimes
  f_i^{(3)} (Q^2), \label{eq:Fnl}\\
 F^{(4)} (x, Q^2) &= x \int_x^1 \frac{dy}{y} \sum_{i = g, q, \bar{q}, h, \bar{h}}
  C_i^{(4)} \left( \frac{x}{y}, \alpha_s^{(4)} (Q^2) \right)
  f_i^{(4)} (y, Q^2)\nonumber\\&=\sum_{i = g, q, \bar{q}, h, \bar{h}}
  C_i^{(4)} \left(\alpha_s^{(4)} (Q^2) \right)\otimes
  f_i^{(4)} (Q^2), \label{eq:Fnf}
\end{align}
in terms of hard coefficient functions $C_i$ and PDFs $f_i$, and it is
understood that, in Eq.~(\ref{eq:fonll}),  in
the 3FS contributions,  PDFs and
$\alpha_s$ are re-expressed perturbatively in terms of their
4FS counterparts. The
structure function $F^{(3,0)} (x,Q^2)$ is the sum of all 
the contributions to the massless-scheme computation which are
already contained in the massive-scheme one. Its subtraction thus avoid
double counting of these contributions, which are contained both in
the four-flavor expression $F^{(4)} (x,Q^2)$, from which they can be
extracted by expanding in powers of the strong coupling, and in the 
three-flavor expression $F^{(3)} (x,Q^2)$, where they correspond
to the sum of all
contributions which do not vanish as the heavy
quark mass tends to zero, namely constants, and collinear logarithmic terms of
the form $\ln Q^2/m^2_h$, which  in  $F^{(4)} (x,Q^2)$ are present as
a consequence of  perturbative
evolution of the PDFs.

The  decomposition Eq.~(\ref{eq:fonll}) holds generically for all structure
functions: $F_2$, $F_1$, $F_3$, neutral current and charged current.
It is useful to decompose the  structure function Eq.~(\ref{eq:fonll})
in a
``heavy'' and ``light'' component
\begin{equation}
\label{eq:fhl}
F (x, Q^2)=F_h (x, Q^2)+F_l (x, Q^2),
\end{equation}
defined respectively as the contribution to $F (x, Q^2)$ which
survives if only the electric (or weak) charge of the heavy quark is non-zero, or
that which survives if the electric and weak charge of the heavy quark
vanishes. The expressions Eq.~(\ref{eq:fonll}) then separately apply
to $F_h$ and $F_l$.

In the remainder of this paper we will
consider the structure function Eq.~(\ref{eq:fonll}) as our basic hard
observable. The generalization to other observables, and specifically
to hadronic processes will in general
require a relabelling of perturbative orders. Indeed, in general, the
perturbative order at which the 3FS and 4FS cross-sections start being
non-zero is
process dependent, and thus so is what one calls leading,
next-to-leading, and so on.

As mentioned, an advantage of the FONLL
method is that the perturbative order at which heavy quark terms are
included in $F^{(3)} (x,Q^2) $ and 
 $F^{(4)} (x,Q^2) $ can be chosen freely. In Ref.~\cite{Forte:2010ta}
(to which we refer for more details)
in particular, three cases were considered explicitly:
FONLL-A, in which $ F^{(3)} (x, Q^2)$ is computed 
 up to order $\alpha_s$, while  $ F^{(4)} (x, Q^2)$ is determined up
 to the next-to-leading log (NLL) level; FONLL-B in which $ F^{(3)}
 (x, Q^2)$ is to order $\alpha_s^2$ and   $ F^{(4)} (x, Q^2)$  is up to NLL;
 and FONLL-C in which $ F^{(3)}
 (x, Q^2)$ is to order $\alpha_s^2$ and   $ F^{(4)} (x, Q^2)$ is up to
 NNLL.

If we wish to include a fitted heavy quark PDF, both  $ F^{(3)}
 (x, Q^2)$ and  $ F^{(4)}
 (x, Q^2)$ must be modified. The modification of the 4FS expression is straightforward. In the absence of a fitted heavy quark
PDFs, the 4FS scheme heavy-quark (and antiquark) PDFs 
are completely determined by perturbative evolution from a vanishing
boundary condition at a  scale of the order of the quark mass. 
For simplicity,   and
in analogy to Ref.~\cite{Forte:2010ta}, in this work we set this scale 
equal to the quark mass itself. Then,
for all $Q^2>m^2_h$, $f^{(4)}_h(x,Q^2)$ and $ f^{(4)}_{\bar h}(x,Q^2)$
satisfy  perturbative evolution with $n_f=4$,
with the boundary condition at $Q^2=m_h^2$ determined by standard VFN
matching to  $f^{(3)}_h(x,m_h^2)=f^{(3)}_{\bar h}(x,m_h^2)=0$.
With a fitted heavy
quark PDF the vanishing condition is relaxed, and $f^{(4)}_h(x,Q_0^2)$ and
$f^{(4)}_{\bar h}(x,Q_0^2)$, with $Q_0\sim m_h$, are just given by some parameterization, 
with parameters to be determined
by comparing to experimental data.

In the presence of a fitted heavy quark component,
the 3FS heavy PDFs
$ f^{(3)}_h(x,m_h^2)$ and $f^{(3)}_{\bar h}(x,m_h^2)$ are thus non-zero, 
and heavy quark  PDFs must then be introduced for
consistency, at all scales. Because in this scheme the heavy quark is
treated as a massive object which decouples from renormalization-group
equations, these PDFs are scale independent.

When a fitted heavy quark PDF is
introduced,  the expression of the 4FS structure functions
Eq.~(\ref{eq:Fnf}) is thus unchanged: the only change is in the boundary condition satisfied by
the heavy quark PDFs when solving the evolution equations. The
expression of the  3FS structure functions
Eq.~(\ref{eq:Fnl}) instead does change,  because
new contributions to the structure function arise, namely
those with a heavy quark in the initial state.
%

\begin{figure}[h]
  \centering
  \includegraphics[width=0.32\textwidth]{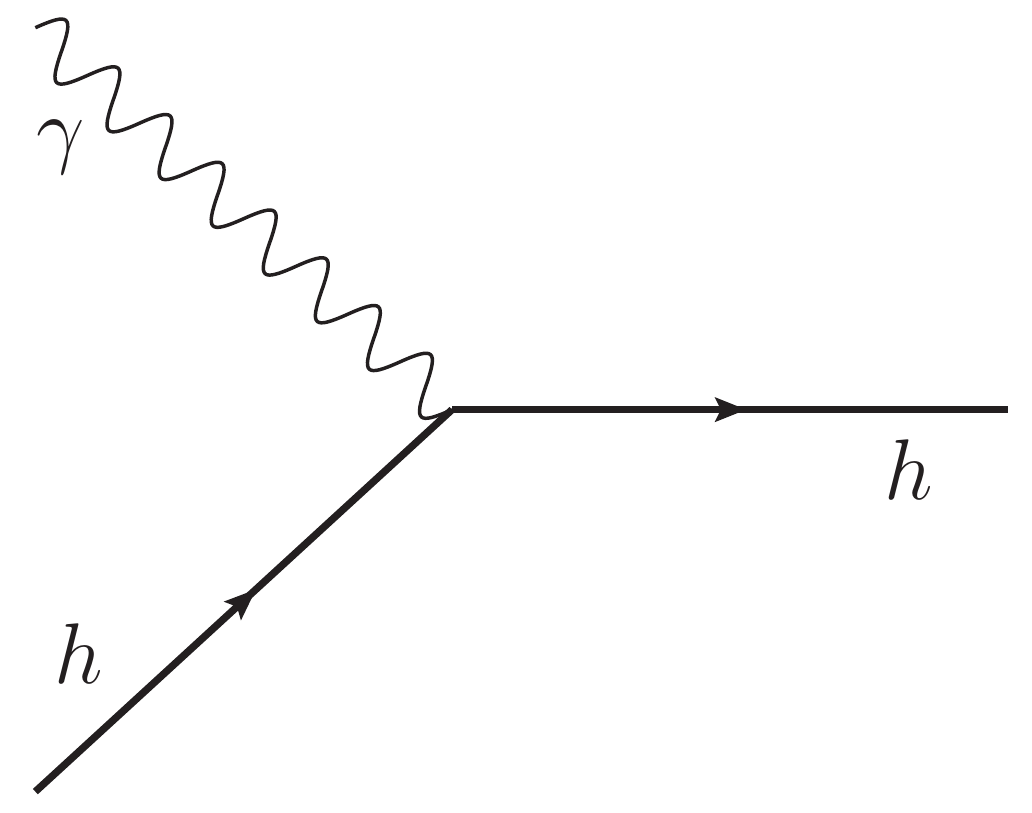}
  \includegraphics[width=0.32\textwidth]{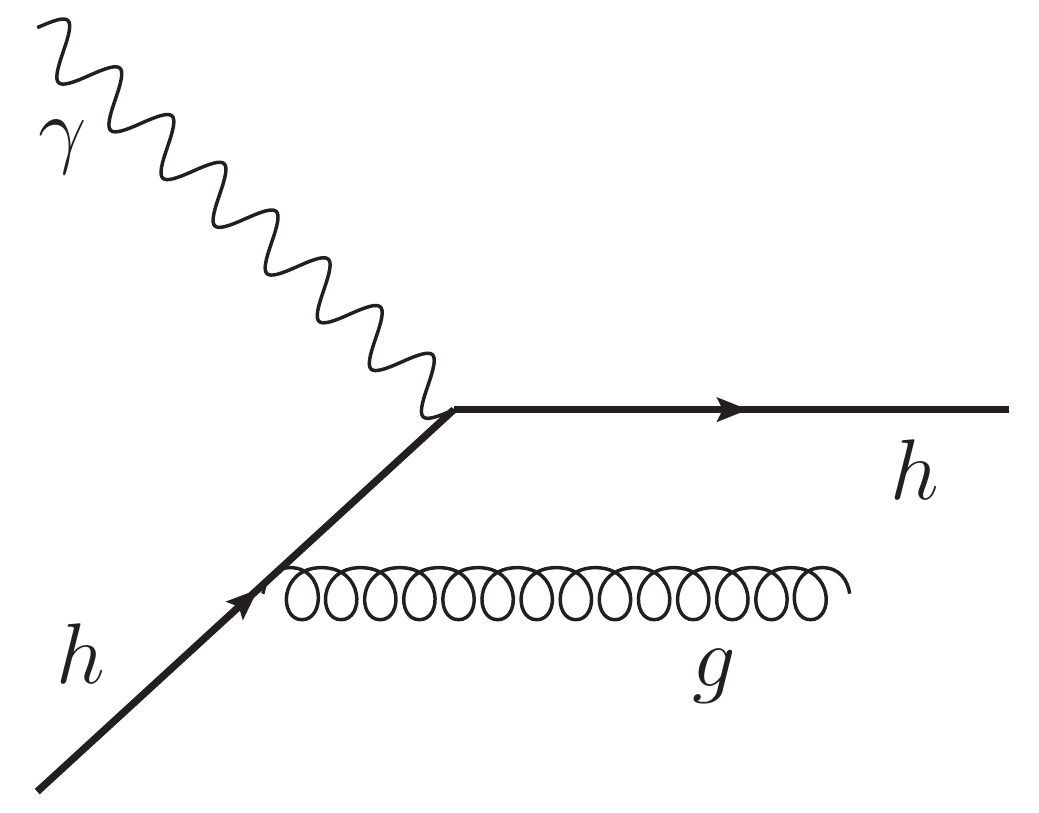}
  \includegraphics[width=0.32\textwidth]{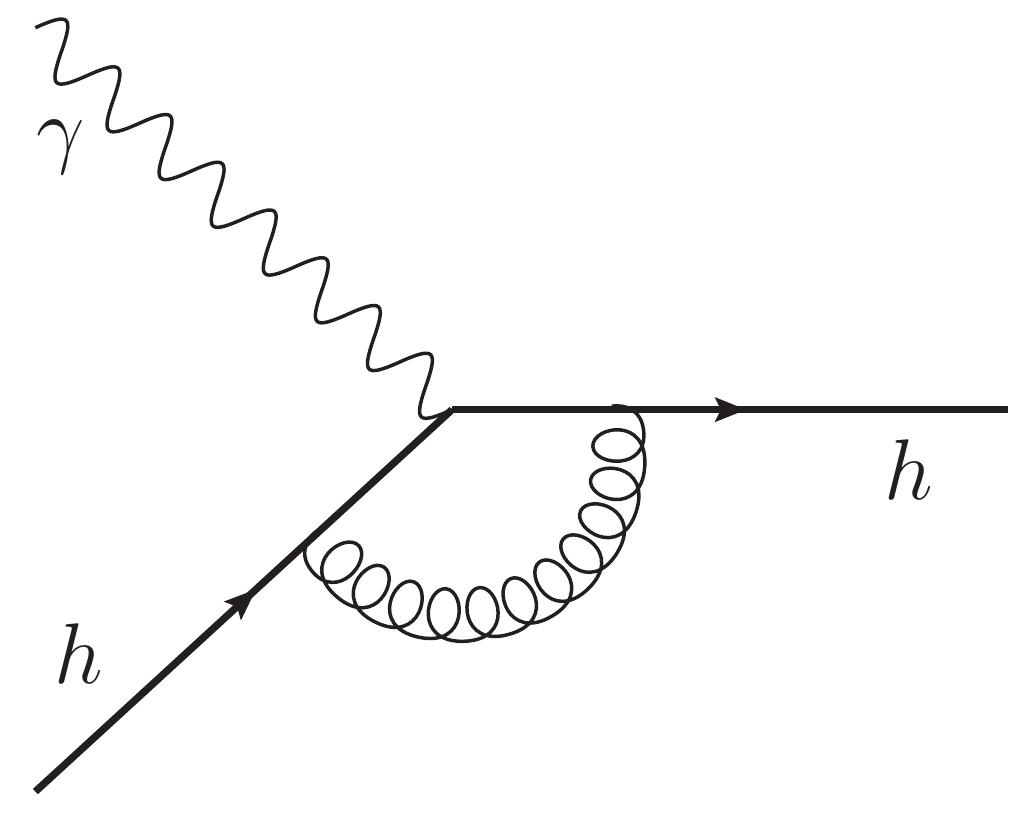}
  \caption{\small Representative Feynman diagrams for the contributions to the heavy $F_h(x,Q^2)$
    structure function induced by a  heavy quark PDF. The fermion line
    represents the heavy quark. From left to right, the LO diagram and
    the NLO real and virtual diagrams are shown.
    }
  \label{fig:feynman}
\end{figure}

Specifically,
decomposing the structure
functions into a heavy and light contribution according to
Eq.~(\ref{eq:fhl}), the heavy structure functions $F_h$ 
now receive a contribution from
$f^{(3)}_h$ and  $ f^{(3)}_{\bar h}$ which starts at $\mathcal O(\alpha_s^0)$
(i.e.\ at the parton-model level). The corresponding coefficient
functions have been computed, both in the neutral- and
charged-current sector, in
Refs.~\cite{Hoffmann:1983ah,Kretzer:1998ju}, up to $\mathcal O(\alpha_s)$.
The corresponding Feynman diagrams are shown in Fig.~\ref{fig:feynman}.
The light structure functions $F_l$ instead receive a contribution
which starts at  $\mathcal O(\alpha_s^2)$.
Because heavy-quark initiated 
massive contributions are only known up to  $\mathcal O(\alpha_s^2)$ corrections,
the highest accuracy which can be achieved at present
in the inclusion of a fitted heavy quark is FONLL-A.

We  thus 
define a correction term $\Delta F(x,Q^2)$, which must be added
to the standard FONLL structure functions $F^{\rm FLNR}(x,Q^2)$ 
of Ref.~\cite{Forte:2010ta} in order to account for the inclusion of
 a fitted
heavy quark PDF: the structure function Eq.~(\ref{eq:fonll}) is thus now given by
\begin{equation}\label{eq:deltadef}
F(x,Q^2)= F^{\rm FLNR}(x,Q^2)+\Delta F(x,Q^2) \, .
\end{equation}
Because the 4FS expressions are unaffected by the correction, 
only $F^{(3)} (x,Q^2)$ and $F^{(3,0)} (x,Q^2)$ contribute to $\Delta
F(x,Q^2)$, and thus  up to $\mathcal{O}(\alpha_s)$ we find 
\begin{equation}\label{eq:fccont}
\Delta F_h(x,Q^2)= \sum_{i=h,\bar h}\left[C_i^{(3)}
\left(\frac{Q^2}{m_h^2}, \alpha_s^{(3)} (Q^2) \right)-
C_i^{(3,0)}\left(\frac{Q^2}{m_h^2}, \alpha_s^{(3)} (Q^2)
\right)\right]\otimes f_i^{(3)}
\end{equation}
(at higher
orders, further terms due to operator mixing would contribute to
the difference).
Note that $f_h^{(3)}$ and $f_{\bar h}^{(3)}$ are scale-independent.

The actual FONLL expression is obtained by re-expressing the coupling
and PDFs in the 3FS contribution to Eq.~(\ref{eq:fonll}) in terms of their 4FS
counterparts. This is done by first,  matching the coupling and PDFs at
some fixed scale, and then, evolving them in the respective
schemes. Matching at the heavy quark mass we have 
\begin{align}
&\alpha_s^{(4)}(m_h^2)=\alpha_s^{(3)}(m_h^2)+\mathcal{O}(\alpha_s^3)\, ,\nonumber \\
  &f^{(4)}_i(m_h^2)=\sum_{j} K_{ij}(m^2_h)\otimes f_j^{(3)} (m_h^2) \, ,\qquad
  i,j=q,\bar{q},g,h,\bar{h}\, .
\label{eq:matching}
\end{align}
The matching functions
$K_{ij}(m_h^2)=\sum_n \alpha_s^n K^{(n)}_{ij}(m_h^2)$ at zeroth order are of course
$K^{(0)}_{ij}=\delta_{ij}$.  
For $i,j=q,\bar{q},g$, they start receiving non-trivial
contributions at
$\mathcal O(\alpha_s^2)$,   accounting for a two-loop normalization mismatch 
between  quark and gluon operators in the
three- and four-flavour schemes (due to the different number of quark
flavours circulating 
in loops) first determined in Ref.~\cite{pdfnnlo}. 
The $K_{hi}(m_h^2)$ functions, with $i=q,\bar{q},g$, start at 
$\mathcal{O}(\alpha_s^2)$: in the absence of a fitted quark
contribution one may actually  express the 4FS heavy quark PDF in terms
of  massless partons, thus avoiding their explicit
use~\cite{Forte:2010ta}.
The $K_{ih}(m_h^2)$ functions are irrelevant in the absence
of intrinsic charm and are discussed here in the context of FONLL 
for the first time: 
$K_{hh}(m_h^2)$  already  receives non-trivial corrections at
 $\mathcal{O}(\alpha_s)$, while $K_{gh}(m_h^2)$ starts at
$\mathcal{O}(\alpha_s)$, and $K_{qh}(m_h^2)$ starts at higher orders.
It follows that in the absence of a fitted charm component, all
matching conditions coincide with the trivial zeroth-order ones up to
and including
$\mathcal{O}(\alpha_s)$ (FONLL-A), while to $\mathcal{O}(\alpha_s^2)$ (FONLL-B or FONLL-C)
knowledge of the $\mathcal{O}(\alpha_s^2)$ contribution to $K_{qq}(m_h^2)$ is
sufficient~\cite{Forte:2010ta}.  In the 
presence of intrinsic charm, they are already non-trivial at
$\mathcal{O}(\alpha_s)$ (FONLL-A), where  
knowledge of the $\mathcal{O}(\alpha_s)$
correction to $K_{hh}(m_h^2)$ is required. Its explicit expression can
be extracted from the known  $\mathcal{O}(\alpha_s)$ massive
coefficient functions of Ref.~\cite{Hoffmann:1983ah,Kretzer:1998ju},
and  is
given in the Appendix (see Eqs.~\eqref{eq:Ks} below). In order to upgrade to
$\mathcal{O}(\alpha_s^2)$ (FONLL-B or FONLL-C), the yet unknown
$\mathcal{O}(\alpha_s^2)$ correction to  $K_{hh}(m_h^2)$  as well as
the (known)  $\mathcal{O}(\alpha_s)$ contribution to
$K_{gh}(m_h^2)$ would also be required.

Evolving both the three-flavour and
four-flavour quantities in the respective schemes one can turn
Eq.~(\ref{eq:matching}) into matching conditions at any scale $Q^2$:
this then defines a matching matrix $K_{ij}(Q^2)$; of course this will
then generate  logarithmic contributions to
all matching functions starting at  $\mathcal{O}(\alpha_s)$.

In particular, the matching condition satisfied by the heavy quark
PDF at a generic scale $Q^2$ is found recalling that in the 3FS the heavy
quark distribution does not evolve: up to $\mathcal O(\alpha_s)$ one then gets
\begin{align}\label{eq:hqmatch}
&f_h^{(3)} =f_h^{(4)}(Q^2) \nonumber\\
&\quad -\alpha_s^{(4)}(Q^2) \left(K^{(1)}_{hh}(m^2_h)+ P^{(0)}_{qq}L\right)\otimes f_h^{(4)}(Q^2)
-\alpha_s^{(4)}(Q^2) L P^{(0)}_{qg}\otimes g^{(4)}(Q^2) +\mathcal O(\alpha_s^2),
\end{align}
where (following Ref.~\cite{Forte:2010ta}) we have defined
$L\equiv\ln\frac{Q^2}{m_h^2}$ and
$P_{ij}^{(0)}(z)$ are the usual leading-order splitting
functions. Note that, whereas  the 3FS heavy PDF $f_h^{(3)}$  is scale
independent, in practice in Eq.~(\ref{eq:hqmatch}) 
$K_{hh}(Q^2)$ is
expanded out perturbatively and only terms up to
$\mathcal{O}(\alpha_s)$ are kept, thereby inducing
a subleading dependence on the scale $Q^2$
when $f_h^{(3)}$ is expressed in terms of the 4FS PDFs.

We can finally get a simple, explicit expression for $\Delta F(x,Q^2)$ up to $\mathcal O(\alpha_s)$ by noting that, because of
standard collinear factorization together with the matching conditions
Eq.~(\ref{eq:matching}), the
3FS coefficient functions in the massless
limit are simply related to the 4FS mass-independent
coefficient functions: 
\begin{align}\label{eq:czmatch}
&C_i^{(3,0),\,0}\left(\frac{Q^2}{m_h^2}\right)=C_i^{(4),\, 0}  ,\\
\label{eq:comatchq}
&C_i^{(3,0),\,1}\left(\frac{Q^2}{m_h^2}\right)=C_i^{(4),\, 1}+
C_i^{(4),\,  0}\otimes \left( K^{(1)}_{hh}(m^2_h)+  P^{(0)}_{qq}
L\right) ,&&(i=h,\>\bar h),
\end{align}
where 
we have defined $C_i=\sum_n C_i^n \alpha_s^n$.
Eq.~(\ref{eq:czmatch}) holds for any $\mathcal O(\alpha_s^0)$
coefficient function and Eq.~(\ref{eq:comatchq}) holds for
$\mathcal O(\alpha_s^1)$ heavy quark coefficient functions.
Explicit expression for the heavy-quark
initiated massive 3FS
coefficient functions are collected in the Appendix, while the
remaining ones can be found in Ref.~\cite{Forte:2010ta}.

Substituting Eqs.~(\ref{eq:hqmatch}-\ref{eq:comatchq}) into
Eq.~(\ref{eq:fccont}) we obtain
\begin{align}
\label{eq:fcfinal}
\Delta F_h(x,Q^2)&=\sum_{i=h,\,\bar h}\Bigg\{\Big[\left(C_i^{(3),\,0}\left(\frac{Q^2}{m_h^2}\right)-
    C_i^{(4),\,0}\right) \nonumber\\
    &\qquad\qquad+\alpha_s^{(4)}(Q^2) 
\left(C_i^{(3),\,1}\left(\frac{Q^2}{m_h^2}\right)-
C_i^{(4),\,1}\right)\Big]\\
&\qquad\qquad
- \alpha_s^{(4)}(Q^2)  C_i^{(3),\,0}\left(\frac{Q^2}{m_h^2}\right)\otimes
\left(K^{(1)}_{hh}(m^2_h)+P^{(0)}_{qq}L\right)\Bigg\}\otimes  f^{(4)}_i(Q^2)\nonumber\\
&\quad - \alpha_s^{(4)}(Q^2) \sum_{i=h,\bar h}\left(C_i^{(3),\,0}\left(\frac{Q^2}{m_h^2}\right)-
C_i^{(4),\,0}\right)\otimes
P^{(0)}_{qg}L\otimes f^{(4)}_g(Q^2)+\mathcal O(\alpha_s^2),\nonumber
\end{align}
where by $\mathcal O(\alpha_s^2)$ we really mean up to subleading terms with
respect to FONLL-A (i.e. $\mathcal O(\alpha_s^2)$ in the 3FS,
and NNLL in the 4FS).

By definition, $\Delta F_h(x,Q^2)$ Eq.~(\ref{eq:fcfinal}) is a contribution to
the ``heavy'' component $F_h$ Eq.~(\ref{eq:fhl}) of the structure
function. We list for completeness the remaining contributions to
$F_h$ in the FONLL-A scheme, as given in Ref.~\cite{Forte:2010ta}:
\begin{align}\label{eq:oldfonll}
F^{\rm FLNR}_h(x,Q^2)&= \sum_{i=h,\,\bar h}
 \left(C_i^{(4),\,0}+\alpha_s^{(4)}(Q^2) C_i^{(4),\,1}\right)\otimes
 f^{(4)}_i(Q^2)+\alpha_s^{(4)}(Q^2) C_g^{(4),\,1}\otimes
 f^{(4)}_g(Q^2)\nonumber\\
&\quad +\alpha_s^{(4)}(Q^2)  \left(C_g^{(3),\,1} \left(\frac{Q^2}{m_h^2}\right)-
C_g^{(4),\,1}- \sum_{i=h,\bar h}C_i^{(4),\,  0}\otimes P_{qg}^{(0)}L \right)\otimes
f^{(4)}_g(Q^2)\nonumber\\
&\quad + \mathcal O(\alpha_s^2)\nonumber\\
&=\sum_{i=h,\,\bar h}
 \left(C_i^{(4),\,0}+\alpha_s^{(4)}(Q^2) C_i^{(4),\,1}\right)\otimes
 f^{(4)}_i(Q^2)\nonumber\\
&\quad+\alpha_s^{(4)}(Q^2)  \left(C_g^{(3),\,1} \left(\frac{Q^2}{m_h^2}\right)
  -\sum_{i=h,\bar h}C_i^{(4),\,  0}\otimes P_{qg}^{(0)}L\right)\otimes
f^{(4)}_g(Q^2)\nonumber\\ &\quad+ \mathcal O(\alpha_s^2).
\end{align}

Substituting Eq.~(\ref{eq:fcfinal}) and
Eq.~(\ref{eq:oldfonll}) in Eq.~(\ref{eq:deltadef}) 
provides the final expression of the heavy structure
functions in the FONLL-A scheme, with the latter providing the
result in the absence of a fitted heavy quark PDF as given in Ref.~\cite{Forte:2010ta}, and the former the
correction due to a non-vanishing heavy quark PDF.
Note that even if the fitted heavy quark PDFs vanishes
(i.e.\ $f^{(4)}_h(x,m^2_h)=f^{(4)}_{\bar h}(x,m^2_h)=0$), the new contribution
$\Delta F_h(x,Q^2)$  Eq.~(\ref{eq:fcfinal}), though  
subleading,  does not vanish when $Q^2>m_h^2$: it only vanishes when
$Q^2=m_h^2$.  This is due to the
fact that, when re-expressing $f^{(3)}_i(Q^2)$ in terms of
$f^{(4)}_i(Q^2)$,  only terms up to $\mathcal O(\alpha_s)$ were kept,
as can be seen from
Eq.~(\ref{eq:hqmatch}).

This means that, even in the absence of a fitted component, our
expressions are not identical to those of Ref.~\cite{Forte:2010ta},
from which they differ by subleading terms. However,  we will
show  below that in the absence of fitted charm
this difference is completely negligible, so that it makes no
difference in practice whether one
uses
Eq.~(\ref{eq:deltadef}), or  Eq.~(\ref{eq:oldfonll}) as in
Ref.~\cite{Forte:2010ta}. 

As well known~\cite{LHhq}, the FONLL-A expression,
as given by Eq.~(\ref{eq:oldfonll}), coincides with the NLO S-ACOT~\cite{Kramer:2000hn}
GM-VFN scheme result. It is  easy to show that FONLL-A as  given by
Eq.~(\ref{eq:deltadef}) coincides with the original
NLO ACOT~\cite{Aivazis:1993pi,Collins:1998rz} scheme. Indeed, note
that once Eq.~(\ref{eq:fcfinal}) and
Eq.~(\ref{eq:oldfonll}) are combined into  Eq.~(\ref{eq:deltadef})
there is a certain amount of cancellation, and one ends up with the
relatively simpler expression for the heavy structure function
\begin{align}
\label{eq:acot}
  &F_{h}(x,Q^2)=\sum_{i=h,\,\bar h}\Bigg\{C_i^{(3),\,0}\left(\frac{Q^2}{m_h^2}\right)
\nonumber \\
&  +\alpha_s^{(4)}(Q^2) 
\left[C_i^{(3),\,1}\left(\frac{Q^2}{m_h^2}\right)-
  C_i^{(3),\,0}\left(\frac{Q^2}{m_h^2}\right)\otimes \left(K^{(1)}_{hh}(m^2_h)+ 
P^{(0)}_{qq}L\right)\right]\Bigg\}\otimes  f^{(4)}_i(Q^2)\nonumber\\
&+\alpha_s^{(4)}(Q^2)  \left[C_g^{(3),\,1} \left(\frac{Q^2}{m_h^2}\right)
- \sum_{i=h,\bar h}C_i^{(3),\,0}\left(\frac{Q^2}{m_h^2}\right)\otimes
P^{(0)}_{qg}L\right]\otimes f^{(4)}_g(Q^2)\nonumber\\
&+\mathcal O(\alpha_s^2).
\end{align}
In plain words, the result reduces to the expression obtained by
combining the PDFs $f_i^{(4)}$, evolved in the 4FS, with
the massive 3FS coefficient functions $C_i^{(3)}$, and
subtracting from the latter the unresummed collinear logarithms. This
coincides with the ACOT result.

An interesting feature of our result Eq.~(\ref{eq:acot}) is the
following. The FONLL expression Eq.~(\ref{eq:fonll}) can be viewed as
the sum of the 3FS expression, and 
a ``difference'' contribution
\begin{equation} \label{eq:diff}
F^{(d)} (x, Q^2) =   F^{(4)} (x, Q^2) - F^{(3,0)} (x,Q^2),
\end{equation}
which is in fact  subleading with
respect to the accuracy of the massive computation: it only contains
logarithmic terms beyond the order of the 3FS result.
If
the new FONLL expression Eq.~(\ref{eq:deltadef}) is adopted 
the difference term $F^{(d)} (x, Q^2)$ vanishes
identically. This is not accidental: it is due to the fact that, when
re-expressing the 3FS PDFs in terms of the 4FS
ones, Eq.~(\ref{eq:hqmatch}), the difference in evolution is only
compensated up to $\mathcal O(\alpha_s)$. The higher-order collinear logs which
would normally contribute to the difference terms are thus also
included in the 3FS contribution, and subtracted
off.

A consequence of this is that the phenomenologically motivated 
modification of the FONLL expression by
subleading terms (akin to the ACOT-$\chi$~\cite{Tung:2001mv}
modification of ACOT) which was considered in
Ref.~\cite{Forte:2010ta} is no longer possible.
This modification had the purpose of leading to $\mathcal O(\alpha_s)$ (FONLL-A)
results which approximate the full  $\mathcal O(\alpha_s^2)$ (FONLL-B)
result~\cite{Nadolsky:2009ge}. It consisted 
of multiplying $F^{(d)} (x,
Q^2)$ Eq.~(\ref{eq:diff})
by a kinematically motivated function of
$m_h^2/Q^2$ which
tends to one in the large $Q^2$ limit: but this term now vanishes, and
thus this modification would have no effect. 
However, this does not appear to be a limitation, as in the presence of a fitted
charm PDF, subleading terms can now be reabsorbed in the initial PDF.

While we have presented so far results only up to $\mathcal O(\alpha_s)$
(FONLL-A) accuracy, our discussion is easily generalized to higher
orders. Indeed, quite in general, the structure function
Eq.~(\ref{eq:fonll}) has the form
\begin{align}\label{eq:FIC-nf}
&  F(x,Q^2)
= \sum_{i,j=g,q, \bar q,h,\bar h}
\left[C_i^{(3)}\left(\frac{Q^2}{m_h^2}\right)-C_i^{(3,0)}\left(\frac{Q^2}{m_h^2}\right)\right]\otimes
K_{ij}^{-1}(Q^2) \otimes f_j^{(4)}(Q^2)\nonumber \\
&\qquad\qquad\qquad +\sum_{i,j=g,q, \bar q,h,\bar h} C_i^{(4)}  \otimes f_i^{(4)}(Q^2),
\end{align}
where $K_{ij}^{-1}(Q^2)$ is the inverse of the matching
matrix Eq.~(\ref{eq:matching}), and it is understood that all
quantities are re-expressed in terms of $\alpha_s^{(4)}$ and then
expanded out to the desired accuracy, with the 4FS PDFs written in
terms of a set of PDFs at a reference scale through perturbative
evolution in the usual way.

The compact form of Eq.~(\ref{eq:FIC-nf}) reveals an
interesting feature:
using the matching conditions
Eq.~(\ref{eq:matching})  evolved up to a  generic scale $Q^2$,
the second and third
terms in Eq.~(\ref{eq:FIC-nf}) cancel, and one ends up with the very simple
expression
\begin{equation}\label{eq:FIC-nfsimp}
  F(x,Q^2)
= \sum_{\substack{i,j=g,q, \bar q,h,\bar h}}
C_i^{(3)}\left(\frac{Q^2}{m_h^2}\right)\otimes
K_{ij}^{-1}(Q^2) \otimes f_j^{(4)}(Q^2).
\end{equation}
This shows explicitly the vanishing of the difference contribution
Eq.~(\ref{eq:diff}), which thus appears to be an all-order feature of
this approach. Higher-order generalizations then simply require the
determination of the matching matrix $K_{ij}^{-1}(Q^2) $, its perturbative
inversion to the desired order, and the re-expansion of $C_i^{(3)}$ in
terms of $\alpha_s^{(4)}$. An all-order proof of
Eq.~(\ref{eq:FIC-nfsimp}) is given in Ref.~\cite{Ball:2015dpa}, where its
implications are discussed in detail (see in particular
Sect.~3.2 of this reference).

\begin{figure}[t]
  \centering
  \includegraphics[page=1,width=0.49\textwidth]{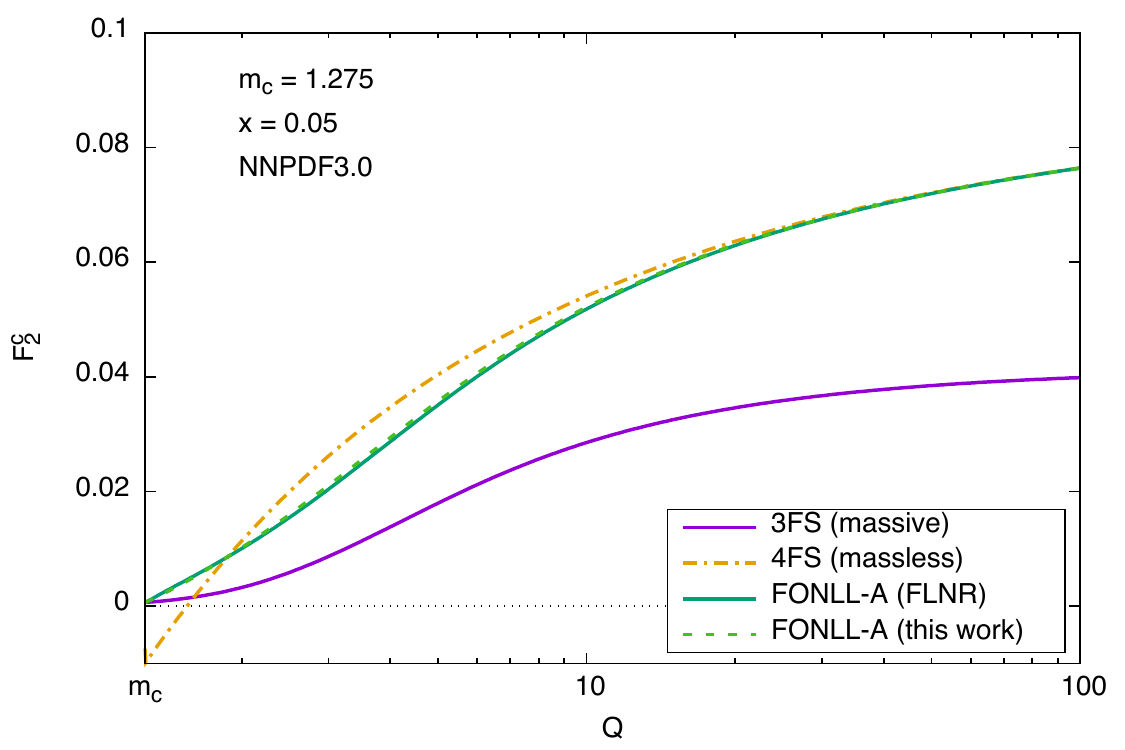}    
  \includegraphics[page=1,width=0.49\textwidth]{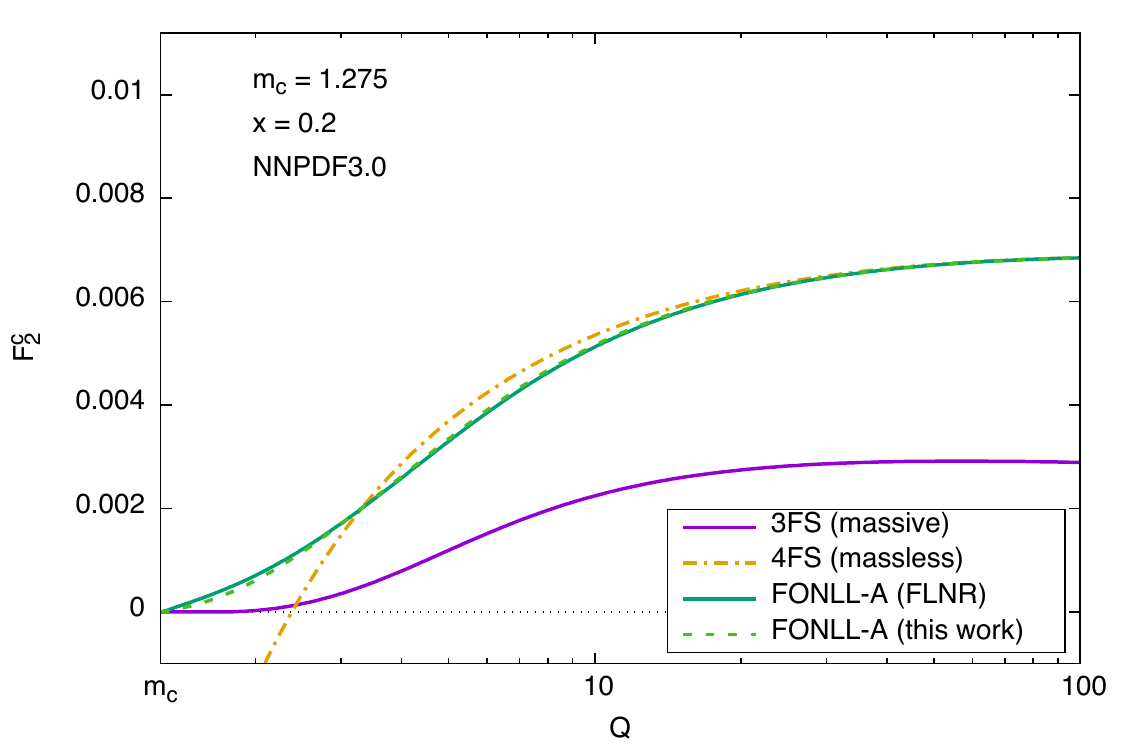}\\
  \includegraphics[page=1,width=0.49\textwidth]{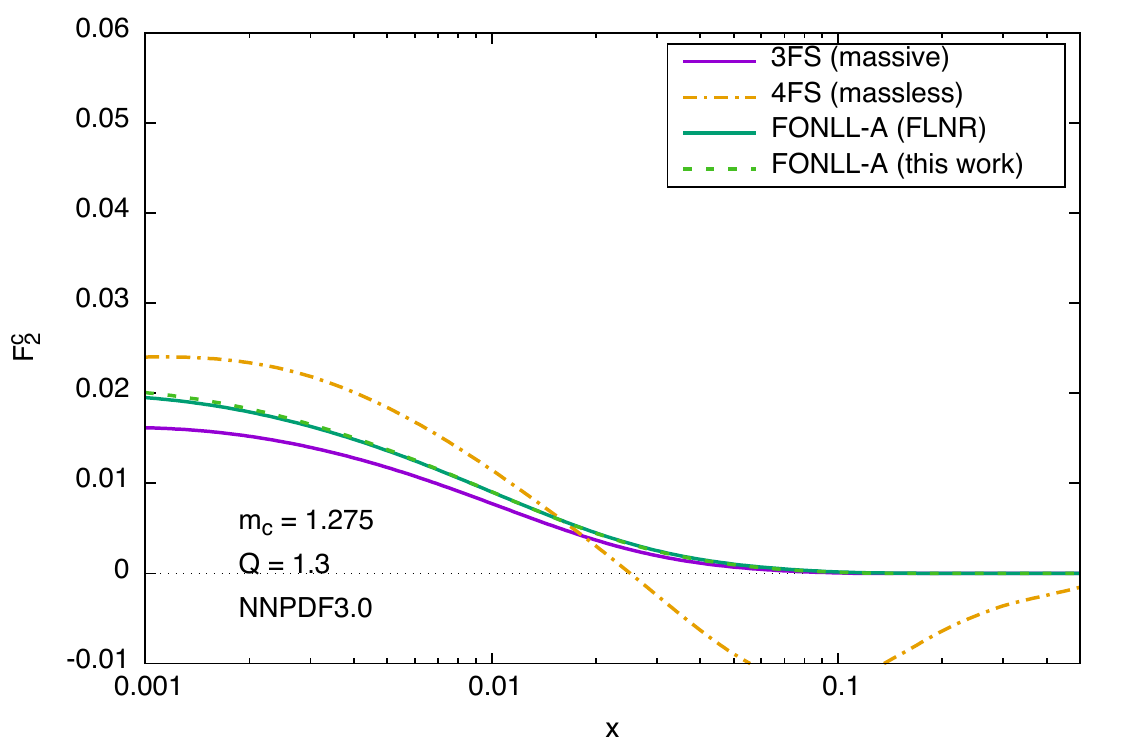}
  \includegraphics[page=1,width=0.49\textwidth]{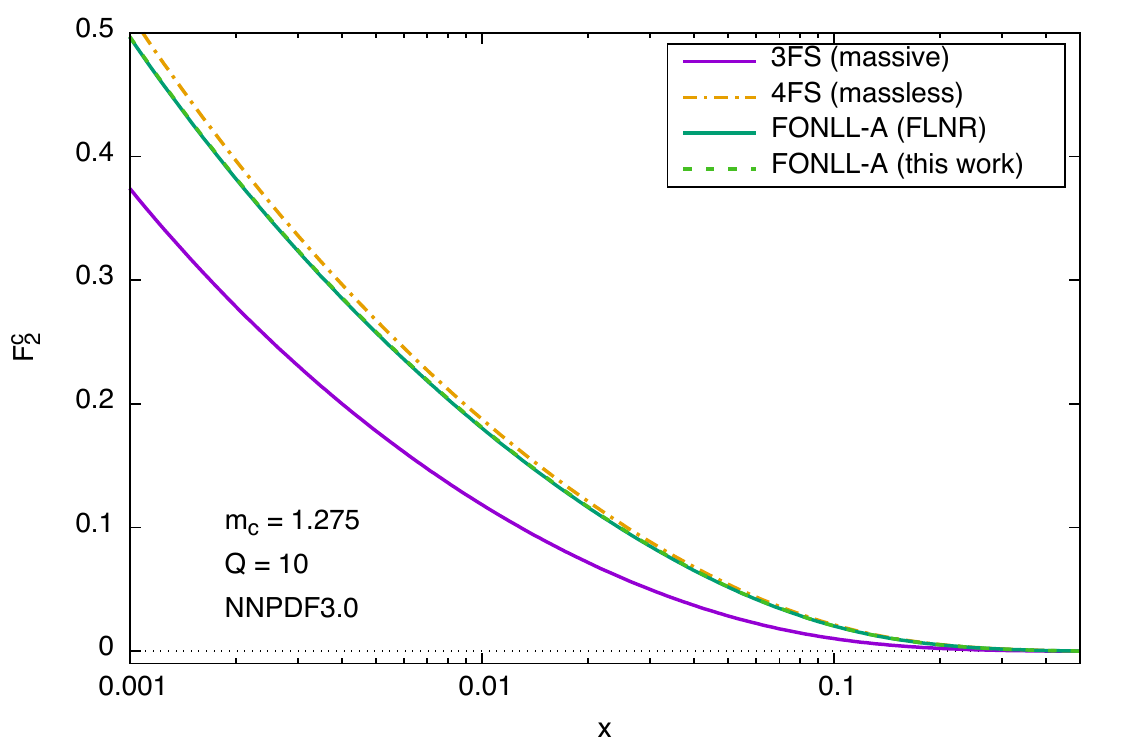}
\caption{\small The charm structure function $F_2^c(x,Q^2)$ in the
    3FS to $\mathcal{O}(\alpha_s)$, in 
the 4FS scheme to NLL, and using the FONLL-A matched scheme,
in the absence of a fitted charm component. The FONLL implementation
of Ref.~\cite{Forte:2010ta} (labelled FLNR) and the implementation of
this paper, which differs from it by subleading terms in the absence
of fitted charm, are both shown. Results are shown as a function of
$Q$ for $x=0.05$ and $x=0.2$, and as a function of $x$ for $Q=1.3$~GeV
and $Q=10$~GeV.
    }
  \label{fig:noIC}
\end{figure}
We finally turn to  a first assessment of 
the phenomenological impact of a possible non-vanishing fitted charm component.
We consider specifically $F_{2}^c(x,Q^2)$, the heavy contribution Eq.~(\ref{eq:fhl}) to the
neutral-current DIS structure function
$F_2(x,Q^2)$.
In the following, results have been obtained  using the
NNPDF3.0 sets~\cite{Ball:2014uwa}, with the corresponding  
value of the charm mass 
$m_c=1.275$ GeV (see Sect.~2.3.4 of Ref.~\cite{Ball:2014uwa}). 
We have generated the results for FONLL-A and FONLL-B structure functions
determined according to the expressions of
Ref.~\cite{Forte:2010ta,LHhq} using \texttt{APFEL}~\cite{Bertone:2013vaa,Carrazza:2014gfa}. 
We have then supplemented them with the extra fitted-charm contributions Eq.~(\ref{eq:fcfinal}),
which was implemented in a new stand-alone public code~\cite{MassiveDIS}.

\begin{figure}[t]
  \centering
  \includegraphics[page=1,width=0.49\textwidth]{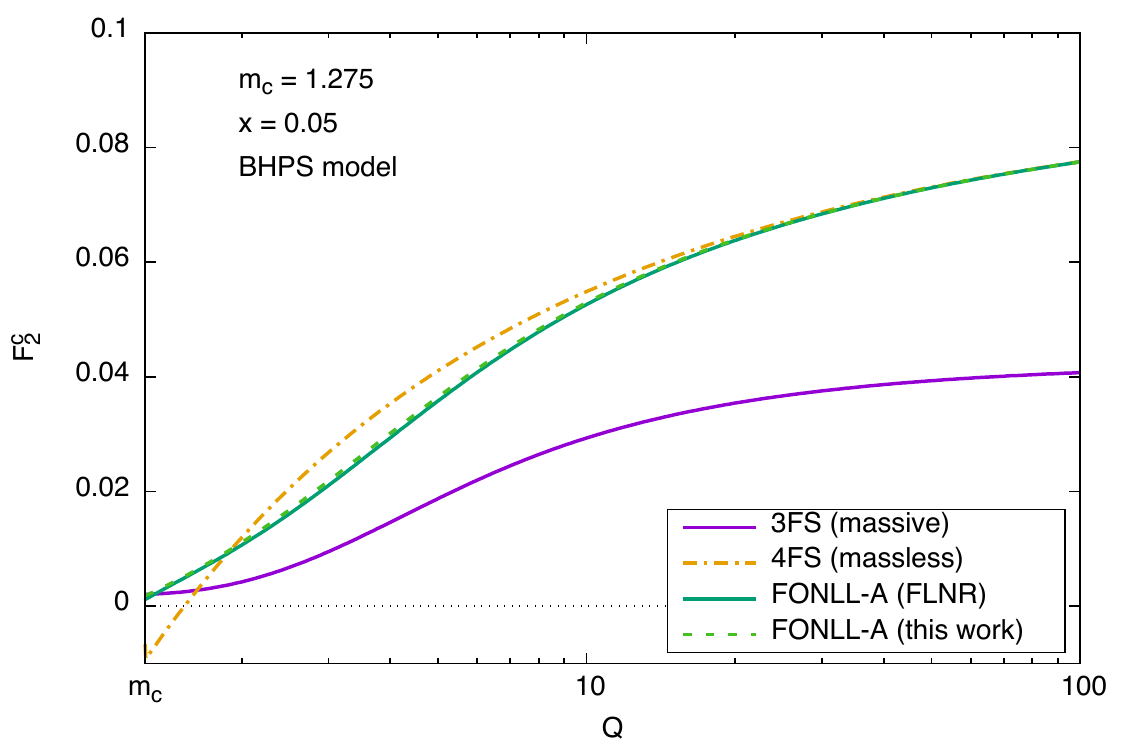}    
  \includegraphics[page=1,width=0.49\textwidth]{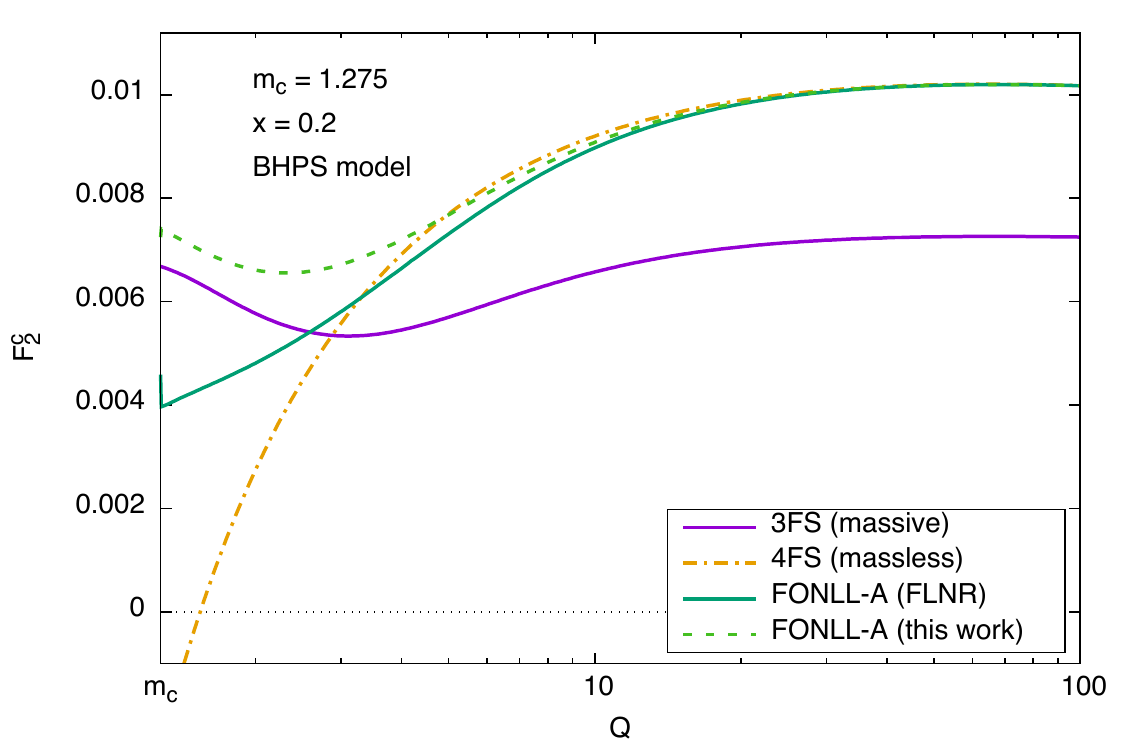}\\
  \includegraphics[page=1,width=0.49\textwidth]{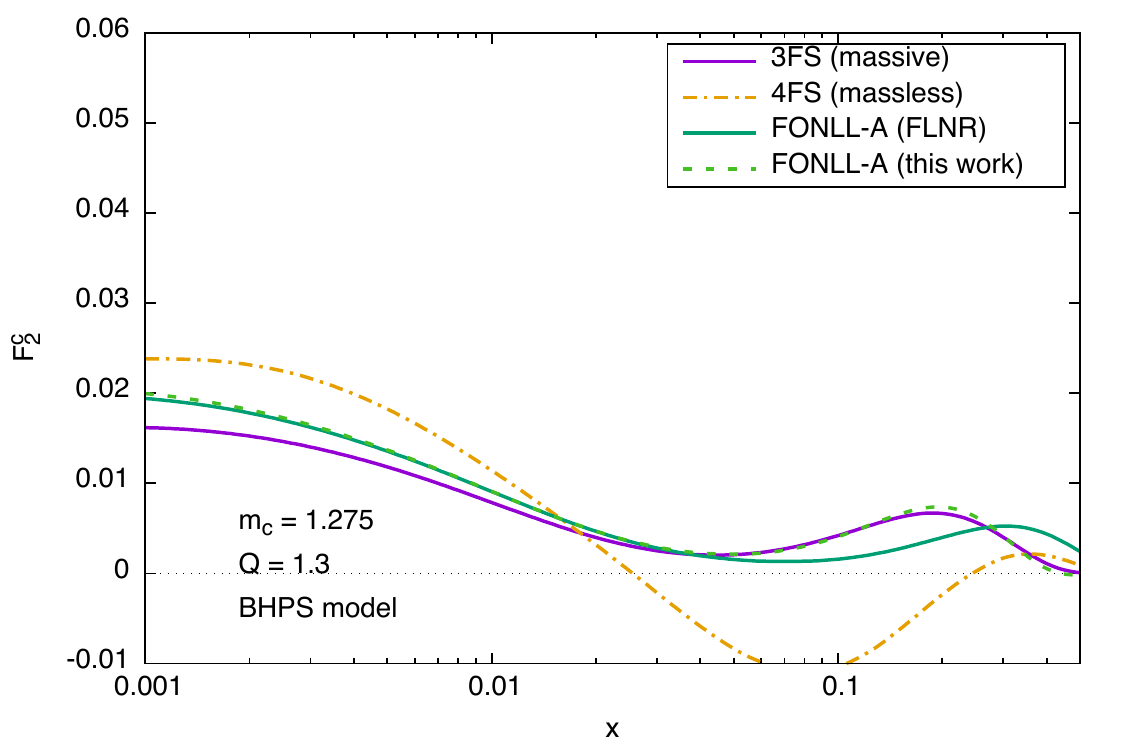}
  \includegraphics[page=1,width=0.49\textwidth]{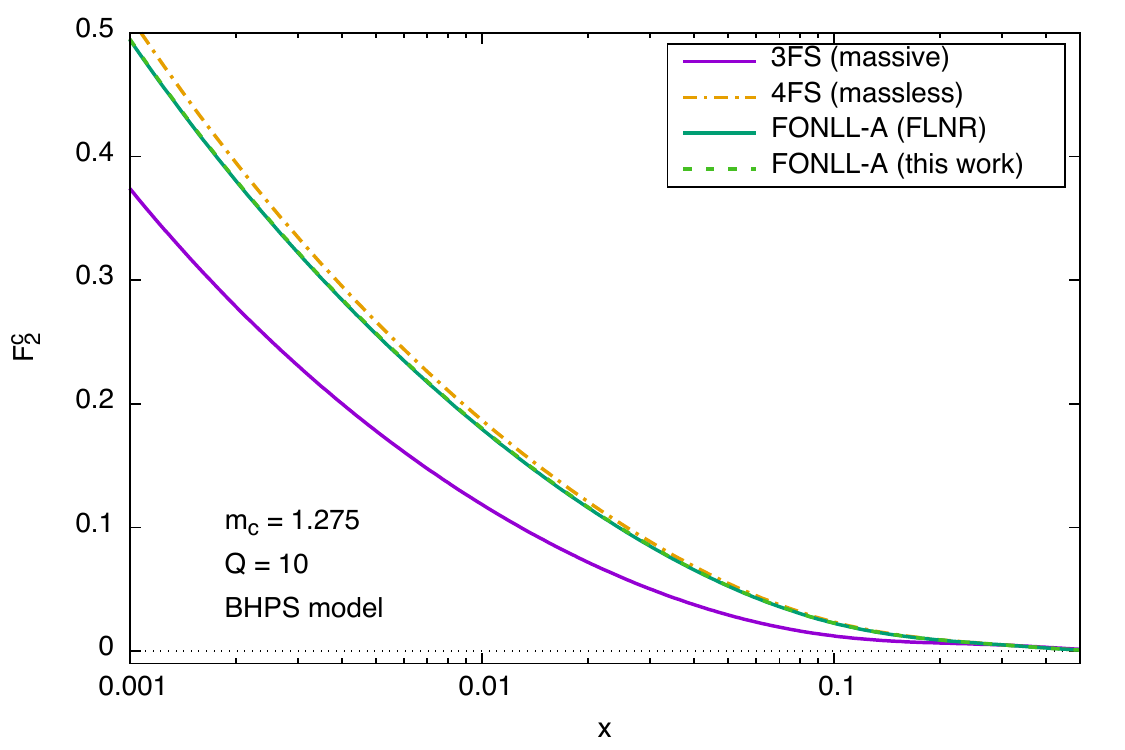}
\caption{\small Same as Fig.~\ref{fig:noIC}, but now including a
  charm component, modeled using the BHPS scenario
  Eq.~(\ref{eq:ICparam}). The FLNR curve, same as in
  Fig.~\ref{fig:noIC}, shows the result in the absence of charm.
    }
  \label{fig:IC5}
\end{figure}
For the sake of a first qualitative assessment, we have generated  a
``fitted'' charm component by assuming two different models
for the charm PDF $f^{(3)}_c(x)=f^{(3)}_{\bar c}(x)$, 
and then matching it to the 4FS expressions
  for $Q^2 \ge m_c^2$. 
We specifically model $f^{(3)}_c(x)=f^{(3)}_{\bar c}(x)$
by using the ``intrinsic charm'' model of Ref.~\cite{Brodsky:1980pb},
which we will refer to as the BHPS model:
\begin{equation}\label{eq:ICparam}
f^{(3)}_c(x)=f^{(3)}_{\bar c}(x)= A\, x^2 [6x(1+x)\ln x +(1-x)(1+10x+x^2)] \, .
\end{equation}
In this model $f^{(3)}_c(x)=f^{(3)}_{\bar c}(x)$ is peaked strongly around $x\simeq 0.2$.
Alternatively, we consider a  scenario, which we refer to as SEA model,
in which the shape of the fitted charm is similar to  that of
all  light quark
sea PDFs, as one would expect if charm was generated by
evolution. For illustrative purposes, we thus take
\be
\label{eq:ICparam2}
f^{(3)}_c(x)=f^{(3)}_{\bar c}(x)= A\, x^{-1.25} (1-x)^3 \, ,
\ee
which has been verified to provide a reasonably good approximation to
the sea PDFs of the  NNPDF3.0 NLO set.

For both scenarios, BHPS, Eq.~(\ref{eq:ICparam}), and SEA, Eq.~(\ref{eq:ICparam2}), we
determine
the value of the normalization
constant $A$, i.e.\ the overall size of the fitted charm contribution, by imposing
that the momentum fraction carried by it is equal to a fixed amount, which we take to be
\begin{equation}
  \label{eq:c_msr}
\langle x \rangle_{c+\bar{c}}(Q_0)  \equiv \int_0^1 dx \ x \lp 
f^{(3)}_c(x)+f^{(3)}_{\bar c}(x)  \rp =5\cdot10^{-3} \, ,
\end{equation}
roughly in line with the phenomenological estimate of  the
CT10IC study~\cite{Dulat:2013hea}.
These starting PDFs have then been combined with the gluon and the
light quark PDFs from the NNPDF3.0, adjusting the gluon
in order to ensure that the
momentum sum rule still holds after accounting for
Eq.~(\ref{eq:c_msr}), and evolved to all scales using
\texttt{APFEL}. Note that in an actual PDF fit, it would be more advantageous to
directly parameterize the heavy quark PDF above threshold, in the 4FS.

\begin{figure}[t]
  \centering
  \includegraphics[page=1,width=0.49\textwidth]{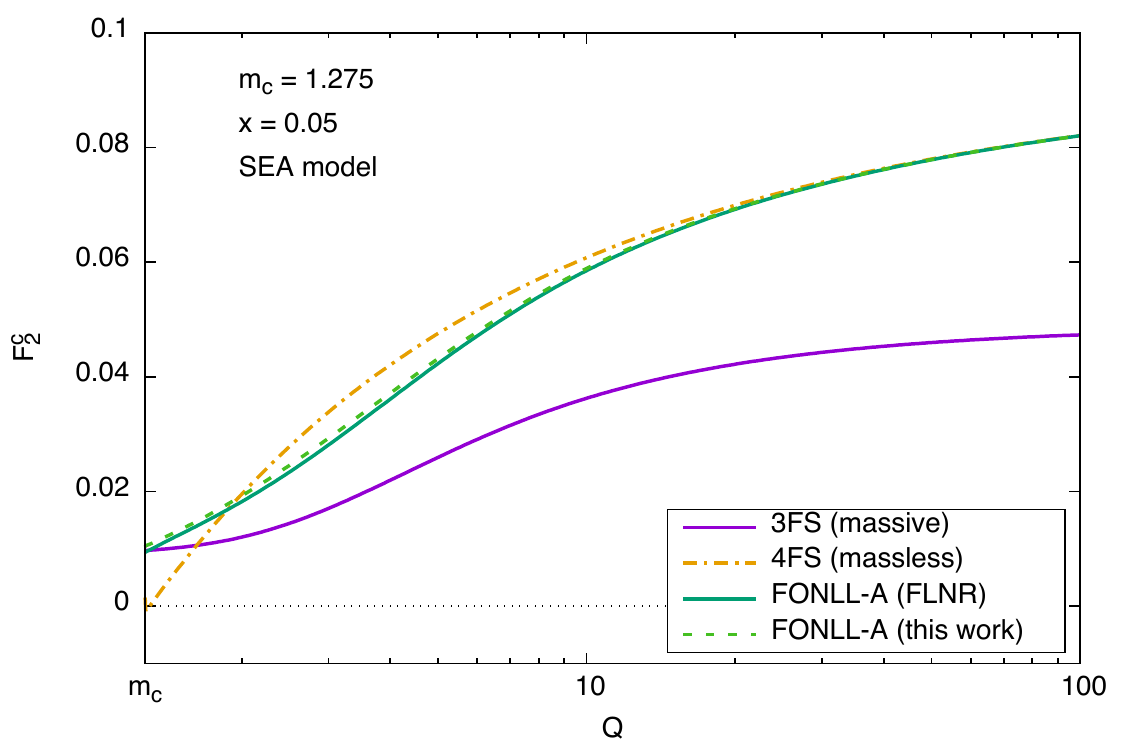}    
  \includegraphics[page=1,width=0.49\textwidth]{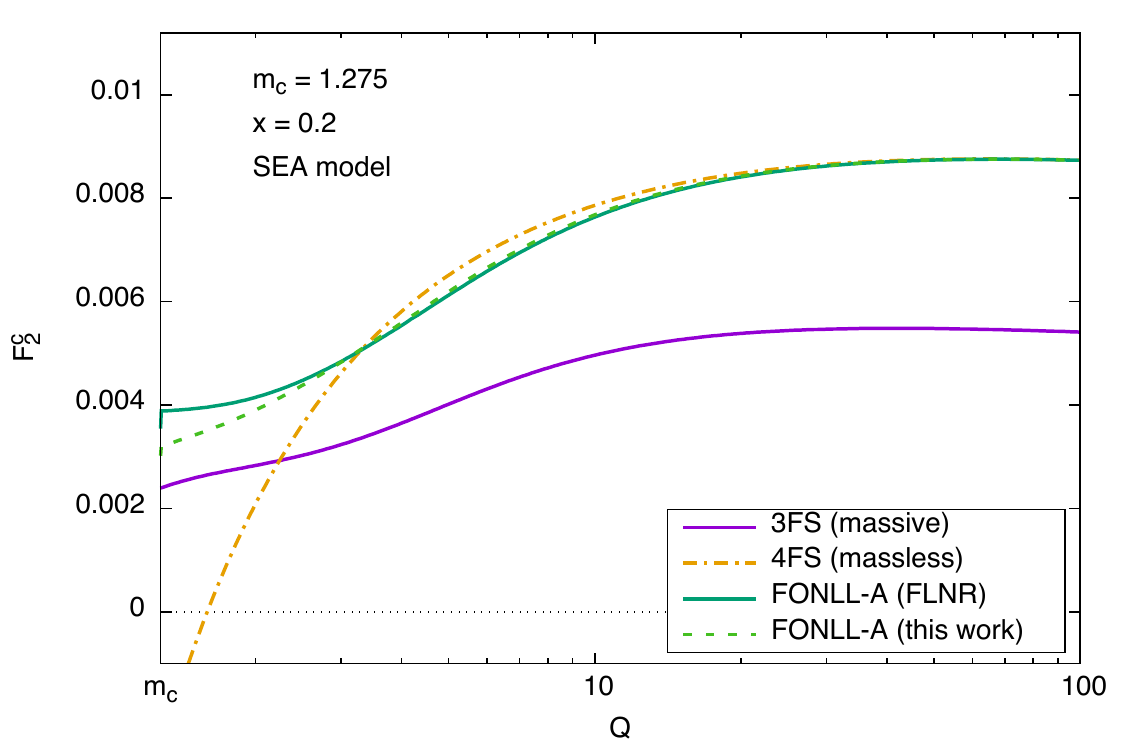}\\
  \includegraphics[page=1,width=0.49\textwidth]{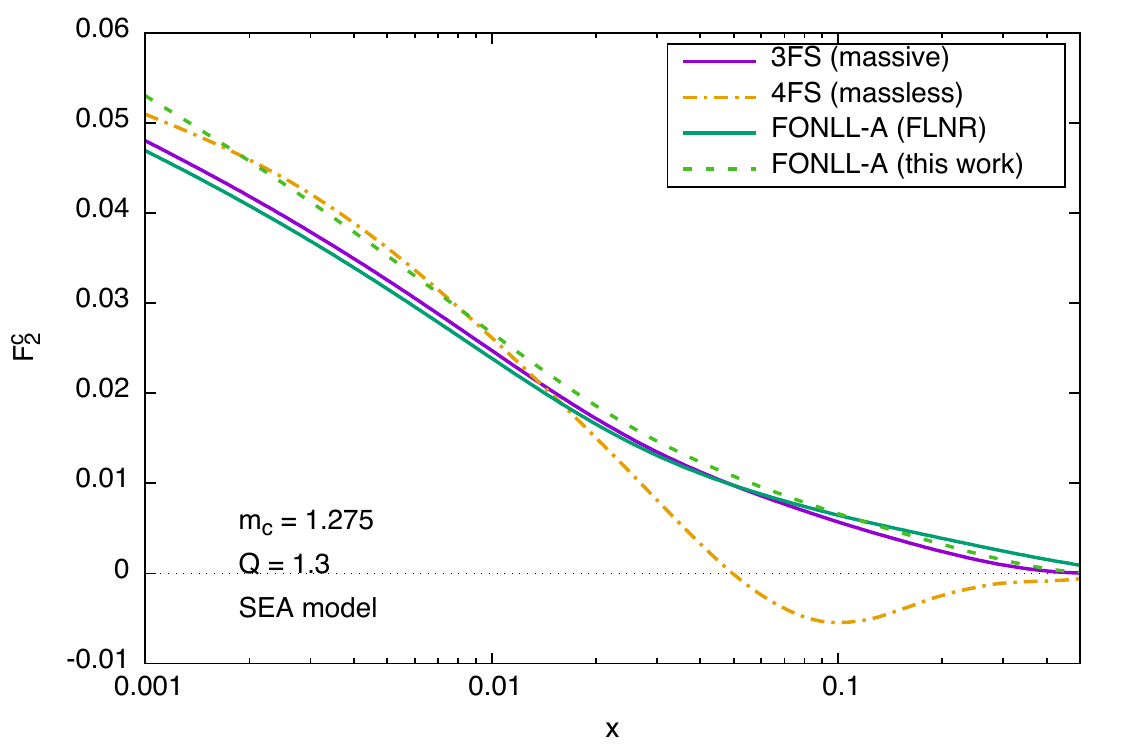}
  \includegraphics[page=1,width=0.49\textwidth]{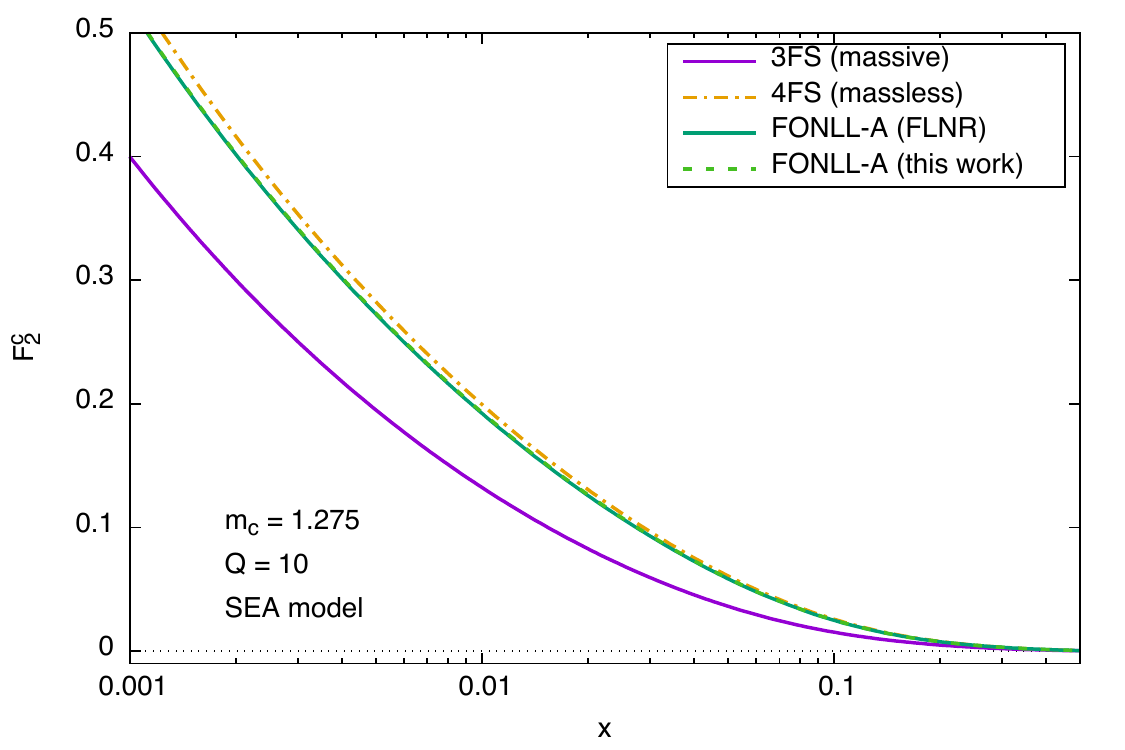}
\caption{\small Same as Fig.~\ref{fig:IC5}, but now with 
 the  charm component modeled using the SEA scenario
  Eq.~(\ref{eq:ICparam2}).
    }
  \label{fig:ICSEA}
\end{figure}
Before turning to these models, we first check that the
 modification of the FONLL scheme of Ref.~\cite{Forte:2010ta},
 Eq.~(\ref{eq:fcfinal}), which is subleading in the absence of a
 fitted charm component, is indeed
negligible for all practical purposes, as mentioned above.
In Fig.~\ref{fig:noIC}, we show, for two
different values of Bjorken $x$ as a function of $Q$ and for two
different values of $Q$ as a function of $x$, the charm structure function
$F_2^c(x,Q^2)$ 
computed to $\mathcal{O}(\alpha_s)$ in the 3FS, Eq.~\eqref{eq:Fnl} (with PDFs
and $\alpha_s$ also in the 3FS), in the 4FS NLL
 Eq.~\eqref{eq:Fnf},
and using the FONLL-A matched scheme.
In the latter case, we both show the original FONLL result
Eq.~(\ref{eq:oldfonll}) and the new form of FONLL presented here,
which includes the extra term 
$\Delta F_h(x,Q^2)$  Eq.~(\ref{eq:fcfinal}).
It is clear that the correction is indeed negligible: this means that
when the fitted charm component vanishes the FONLL result of
Ref.~\cite{Forte:2010ta} is reproduced.

\begin{figure}[t]
  \centering
  \includegraphics[page=1,width=0.49\textwidth]{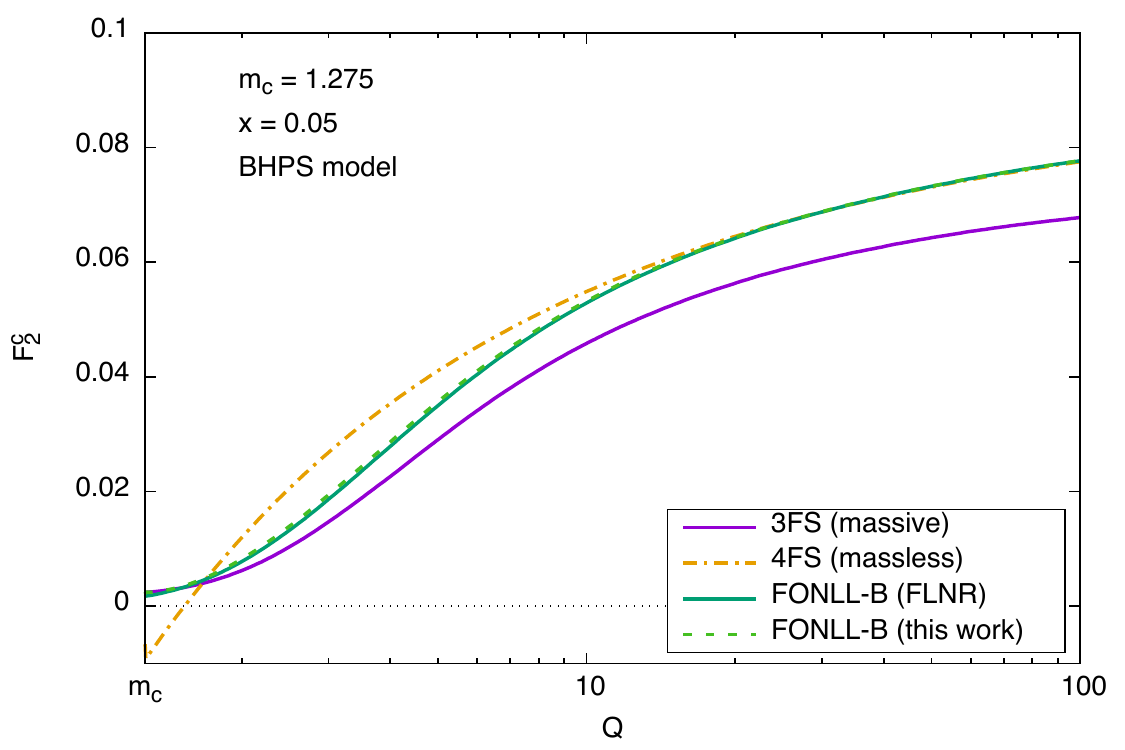}    
  \includegraphics[page=1,width=0.49\textwidth]{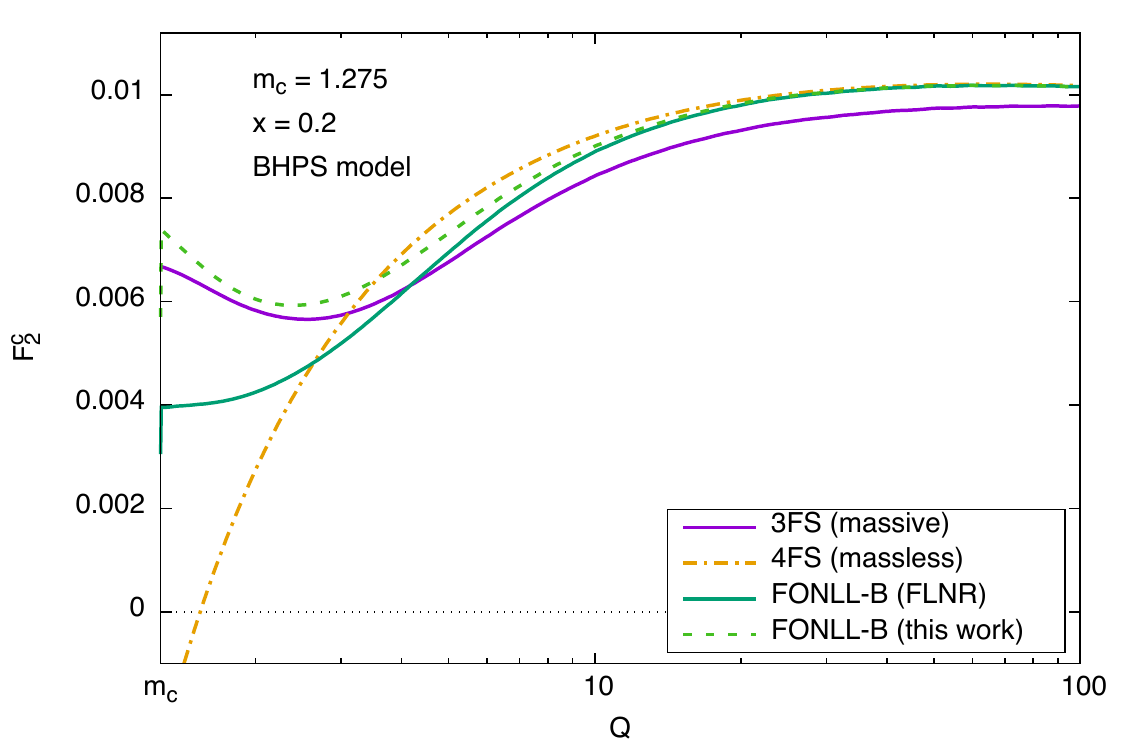}\\
  \includegraphics[page=1,width=0.49\textwidth]{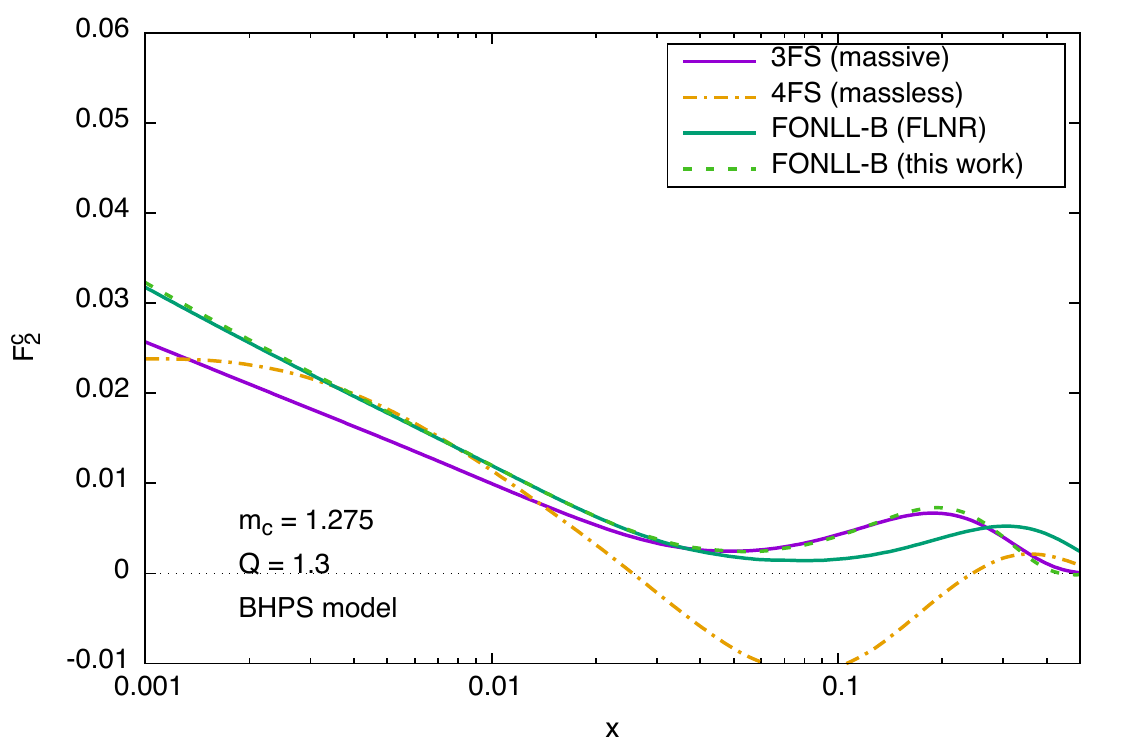}
  \includegraphics[page=1,width=0.49\textwidth]{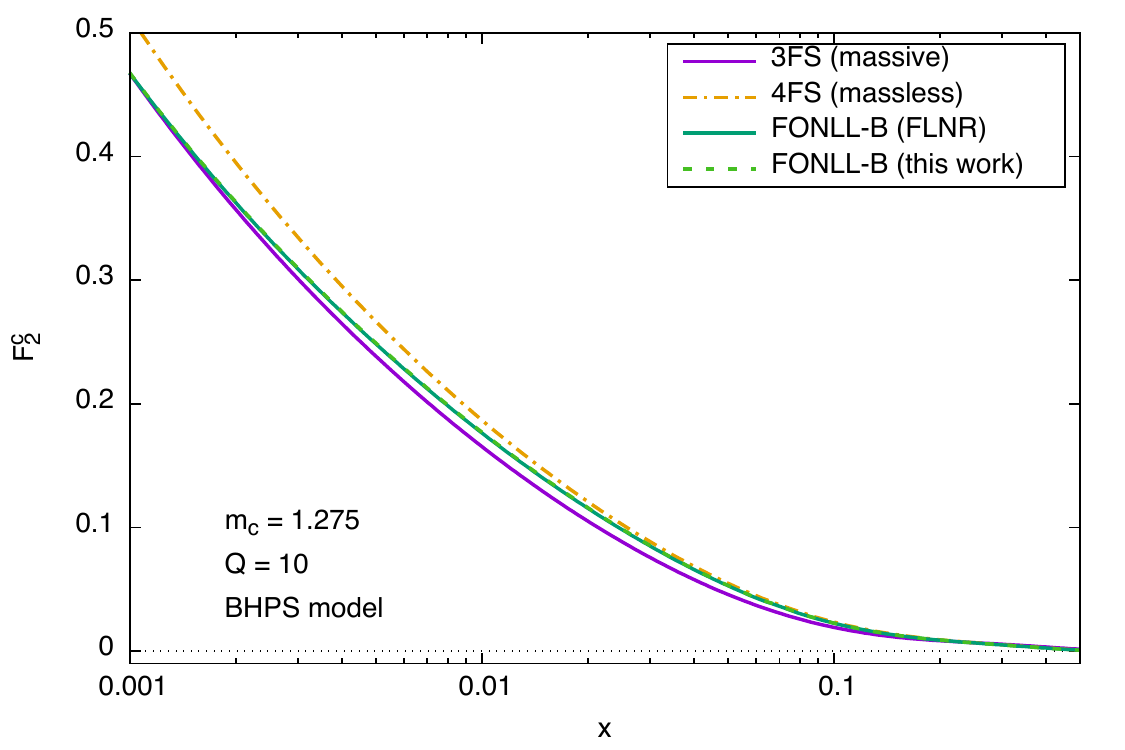}
\caption{\small Same as Fig.~\ref{fig:IC5}, but now using with an
  approximate 
FONLL-B matched scheme (see text). 
    }
  \label{fig:IC5B}
\end{figure}

We now include a fitted charm component. Results are shown 
in Fig.~\ref{fig:IC5} for the
BHPS scenario Eq.~(\ref{eq:ICparam}), and in Fig.~\ref{fig:ICSEA} for
the SEA scenario Eq.~(\ref{eq:ICparam2}).
For the BHPS scenario, the charm contribution is only significant for large
$x\gtrsim 0.08$, and peaks around $x\sim0.2$, while for the SEA model
it provides a non-negligible correction at small $x$. In both cases,
the ``fitted'' charm component is significant just above threshold,
but already for $Q\sim 10$~GeV it is overwhelmed by the perturbatively
generated component, and the results of Ref.~\cite{Forte:2010ta}
are reproduced.  This suggests that a possible intrinsics component
would be most easily revealed in precise measurements close to the
charm threshold, with only relatively minor effects on LHC processes.

The FONLL-A scheme is the only one for which a fitted charm PDF can be
consistently included, until the  $\mathcal{O}\lp \alpha_s^2\rp$
charm-initiated massive coefficient functions are computed.
However, FONLL-B and FONLL-C provide a rather more accurate description of the low-$Q^2$
charm structure functions, thanks to the inclusion of  $\mathcal{O}\lp \alpha_s^2\rp$
gluon and light-quark initiated terms.
Therefore, a practical compromise could be to assume that the unknown
$\mathcal{O}\lp \alpha_s^2\rp$
charm-initiated massive coefficient functions is in practice negligible,
and simply add the same correction, Eq.~(\ref{eq:fcfinal}), to the standard
FONLL-B result. 
This is likely to be a rather good approximation, given that, as
Figs.~\ref{fig:IC5}-\ref{fig:ICSEA} show, the contribution of the
``fitted'' charm component is actually quite small in all reasonable scenarios.
As a final check, we have thus recomputed predictions in the various
scenarios in this approximate FONLL-B scheme. Results with the BHPS
model are shown in
Fig.~\ref{fig:IC5B}: it is seen that the previous conclusions are
qualitatively unchanged.

In summary, in this work we have
generalized the FONLL GM-VFN
scheme to account for the possibility
that the heavy quark PDF can be fitted from the data, rather than
being generated perturbatively.
The next step  will be to use the calculations presented here
in a global analysis in the NNPDF
framework~\cite{DelDebbio:2007ee,Ball:2008by,Ball:2009mk,
  Ball:2010de,Ball:2012cx}, with the goal of determining
the charm PDF from the data.
This, also thanks to the unbiased NNPDF
methodology, will remove the need to resort to ad-hoc modelling,
and it will allow us to settle quantitatively
a  question
that has been left open
for more than 30 years: is it possible to unambiguously
determine the charm content
of the proton? 
This  will also allow  us to remove any possible bias induced by the
hypothesis that charm vanishes at some more or less arbitrary scale,
and explore the possible  implications of this assumption for
precision phenomenology at the LHC, such as for example to the
determination of the heavy quark masses.

\section*{Acknowledgements}

S.F. thanks Matteo Cacciari for useful discussions, and for hospitality
at LPTHE, Universit\'e Paris VI, where part of this work was done,
supported by a Lagrange award.
V.~B. is supported in part by the ERC grant 291377,
``LHCtheory".
R.~D.~B. and P.~G.~M. are funded by an STFC Consolidated Grant
ST/J000329/1.
S.~F. is  supported in part by an Italian PRIN2010 grant,
    by a
    European Investment Bank  EIBURS grant, and by the European Commission
    through
    the HiggsTools Initial Training
    Network PITN-GA-2012-316704.
    J.~R. is supported by an STFC Rutherford Fellowship and Grant ST/K005227/1 and
    ST/M003787/1.
M.~B., J.~R., L.~R. and V.~B. are
supported by an European Research Council
Starting Grant ``PDF4BSM".
%

\appendix

\section*{Appendix}
\label{app:M}

We collect the explicit expressions for the
additional matching conditions and coefficient functions
that are required to generalize the FONLL scheme to the case in which  a fitted
charm PDF is included, up to FONLL-A accuracy. All other matching and
coefficient functions were given in the Appendix to Ref.~\cite{Forte:2010ta}.

The new matching conditions involve the $K_{ih}$ entries of the
matching matrix Eq.~(\ref{eq:matching}). Up to $\mathcal{O}(\alpha_s)$, only 
 $K_{qh}$ and $K_{hh}$ receive non-trivial contributions, both of which
we give for completeness, even though to FONLL-A accuracy only
$K_{hh}$ is needed:
\begin{subequations}
  \label{eq:Ks}
\bea
  K_{hh} \left(Q^2\right) &= K_{\bar{h}\bar{h}} \left(Q^2\right)=1 + \as \plus{\bar{P}_{qq}^{(0)}(z)\(\ln\frac{Q^2}{m_h^2(1-z)^2}-1\) } +\Ord(\as^2) \, ,\\
  K_{gh}  \left(Q^2\right) &= K_{g\bar{h}}  \left(Q^2\right)=\as P_{gq}^{(0)}(z) \(\ln\frac{Q^2}{m_h^2z^2}-1\) +\Ord(\as^2) \, .
\eea
\end{subequations}
In Eq.~(\ref{eq:Ks}) 
\bea
\bar{P}_{qq}^{(0)}(z)= \frac{C_F}{2\pi}\frac{1+z^2}{1-z},\qquad
P_{gq}^{(0)}(z)= \frac{C_F}{2\pi}\frac{1+(1-z)^2}{z} \, ,
\eea
where $P_{qq}^{(0)}(z)=[\bar{P}_{qq}^{(0)}(z)]_+$.  The
expressions  Eq.~(\ref{eq:Ks}) can be obtained by combining the
matching functions Eq.~(\ref{eq:matching})  at $Q=m_h$ with standard
perturbative evolution:
\begin{equation}
K^{(1)}_{ij} \left(Q^2\right)= K^{(1)}_{ij} \left(m_h^2\right) +L
P^{(0)}_{ij},
\end{equation}
where $P_{ij}^{(0)}$
denote the usual leading-order splitting functions, with
$P_{hh}^{(0)}=P_{qq}^{(0)}$ and $P_{gh}^{(0)}=P_{gq}^{(0)}$.

In order to write the charm-initiated massive coefficient
functions up to $\mathcal O(\as) $,
we introduce a number of useful definitions:
\bea
\lambda = m_c^2/Q^2 \, , \qquad
\chi  = \frac{ x(1 + \sqrt{1+4\lambda})}{2}\, .\label{eq:kindefs}
\eea
The contribution 
from the subprocess $\gamma^* c \rightarrow c$ to the charm structure function
$F_{2,c}^{(3)} (x,Q^2)$ in the massive calculation can be written as
\bea
\label{eq:deltaF2c}F_{2,c}^{(3)}\big|_{f_c}= x \int_\chi^1 \frac{d\xi}{\xi} C_{2,c}^{(3)}\( \xi,\smallfrac{Q^2}{m_c^2}\) \(f_c^{(3)}\(\smallfrac{\chi}{\xi},Q^2\)+f_{\bar{c}}^{(3)}\(\smallfrac{\chi}{\xi},Q^2\)\)\, ,
\eea
where the $\mathcal{O}\lp \alpha_s^0\rp$ and $\mathcal{O}\lp \alpha_s\rp$
coefficient functions
\bea
C_{2,c}^{(3)}\(\xi,\frac{Q^2}{m_c^2}\) = C_{2,c}^{(3),0} \(\xi,\frac{Q^2}{m_c^2}\) + 
\alpha_s C_{2,c}^{(3),1} \(\xi,\frac{Q^2}{m_c^2}\) + \mathcal O(\as^2),
\eea
have been computed in Refs.~\cite{Hoffmann:1983ah,Kretzer:1998ju}.
Note that the lower limit of the convolution integral in Eq.~(\ref{eq:deltaF2c}) is
$\chi$, hence a change of variable is needed in order to recover the
form Eq.~(\ref{eq:Fnl}) of the convolution.
The complete structure function $F_{2,c}^{(3)} (x,Q^2)$ in the massive scheme is
constructed by adding Eq.~(\ref{eq:deltaF2c}) to the corresponding gluon- and light-quark initiated contributions.

At $\mathcal{O}\lp \alpha_s^0\rp$ the massive coefficient function
for the charm-initiated contribution reads
\begin{align}
  C_{2,c}^{(3),0}\left( \xi,\smallfrac{Q^2}{m_c^2}\right) &= e^2_c \sqrt{1+4\lambda}
  \delta(1-\xi)\, ,\label{eq:C2c0}
\end{align}
whose massless limit, $C_{2,c}^{(3,0),0}$, is given by
\begin{align}
C_{2,c}^{(3,0),0}\left( \xi,\smallfrac{Q^2}{m_c^2}\right) &= e^2_c \delta(1-\xi) \, ,
\end{align}
which of course coincides with  the leading-order massless quark  coefficient function.

At the next order, $\mathcal{O}\lp \alpha_s\rp$, it
is possible to express the massive coefficient function
for the charm-initiated contribution from
Ref.~\cite{Kretzer:1998ju} as follows:
\bea
\qquad&C_{2,c}^{(3),1}\left( \xi,\smallfrac{Q^2}{m_c^2}\right)  
=
 \frac{2}{3}\frac{e^2_c}{\pi}\delta(1-\xi) \sqrt{1+4\lambda}
\Big\{ 4 \ln \lambda - 2 + \sqrt{1+4 \lambda}\tilde L-2 \ln(1+4\lambda)\nonumber\\
&+\frac{1+2 \lambda}{\sqrt{1+4\lambda}} [3 \tilde L^2 + 4\tilde L + 4 \Li_2(-d/a)+ 2\tilde L \ln \lambda - 2\tilde L \ln (1 + 4 \lambda) + 2 \Li_2 (d^2/a^2)] \Big\}
\nonumber\\&\frac{1}{3}\frac{e^2_c}{\pi}\frac{1}{(1-\xi)_+}\frac{(1-\xi)}{\Delta'^2
\xi^4 \hat s_1}\Big\{\frac{1}{2\hat s^2\xi}
   \nonumber
   \\&\Big\{(1-\xi)^2(1-2\xi-6\xi^2-6\xi^6-2\xi^7+\xi^8)
   \nonumber\\&+\lambda (1-\xi)^2\left(5+12\xi-115\xi^2-20\xi^3-4\xi^4-20\xi^5-115\xi^6+12\xi^7+5\xi^8\right)
   \nonumber\\&+\lambda^2 \left(5\xi^{10}+64 \xi^9-361 \xi^8-8 \xi^7+828 \xi^6-1072 \xi^5+828 \xi^4-8 \xi^3- 361\xi^2+64\xi+5\right)
   \nonumber\\&+4\xi\lambda^3 \left(9+24\xi-202\xi^2-112\xi^3+530\xi^4-112\xi^5-202\xi^6+24\xi^7+9\xi^8\right)
   \nonumber\\&+32\lambda^4 \left(\xi^7-10\xi^5+ \xi^3\right)
   \nonumber\\&+\sqrt{4 \lambda+1} \Big[(1-\xi)^3(1+\xi)(1-2\xi-5\xi^2-2\xi^3-5\xi^4-2\xi^5+\xi^6)
   \nonumber\\&+\lambda\left(-3 \xi^{10}-10 \xi^9+132 \xi^8-202\xi^7+79 \xi^6-79 \xi^4+202 \xi^3-132 \xi^2+10 \xi+3\right)
   \nonumber\\&+\lambda^2 (1-\xi)(1+\xi)\left(1+36\xi-98\xi^2-376\xi^3+730\xi^4-376\xi^5-98\xi^6+36\xi^7+\xi^8\right)
   \nonumber\\&+16\lambda^3(1-\xi)\xi^2(1+\xi) \left(2+\xi-26\xi^2+\xi^3+2\xi^4\right)\Big]\Big\}
   \nonumber\\&+\frac{\hat{L}}{\Delta'}
   \Big[-(1+\xi^2)^2(1-\xi^2+\xi^4)-4\lambda \left(1-\xi+6\xi^2-2\xi^4+6\xi^6-\xi^7+\xi^8\right)
   \nonumber\\&-2\lambda^2 \left(1-6\xi+18\xi^2+46\xi^3-54\xi^4+46\xi^5+18\xi^6-6\xi^7+\xi^8\right)
  +16\lambda^3 \left(\xi^2-10\xi^4+\xi^6\right)
   \nonumber\\&+\sqrt{4 \lambda+1} \Big(\xi^8+\xi^6-\xi^2-1+2\lambda \left(-1+2\xi-11\xi^2+11\xi^6-2\xi^7+\xi^8\right)
   \nonumber\\&+4\lambda^2(1-\xi)\xi(1+\xi) \left(1+2\xi-22\xi^2+2\xi^3+\xi^4\right)\Big)
   \Big]\Big\} \, ,
\eea
where we have introduced the following definitions
\bea
d &= \frac{\sqrt{1+4\lambda}-1}{2}, \qquad a = d+1\\
\hat{s}_1&=\frac{(1-\xi)}{\xi}(a+d\xi),\qquad
\hat{s}=\hat s_1-\lambda,\\
\Delta'&=\frac{1}{\xi}\sqrt{\lambda(1+\xi^2)^2+a+d\xi^4)}\\
\tilde L&=\ln\left(\frac{a}{d}\right),\qquad \hat{L}=\ln\left(\frac{1+2\lambda+\hat s_1-\Delta'}{1+2\lambda+\hat s_1+\Delta'}\right)
\eea
and
\bea
\Li_2 (x) = - \int_0^x dz \frac{\ln(1-z)}{z}.
\eea
The massless limit of this coefficient function is
\bea
C_{2,c}^{(3,0),1}\left( \xi,\smallfrac{Q^2}{m_c^2}\right)
&= \frac{e_c^2}{3\pi} \Big\{ \frac{(1-2\xi-6\xi^2)}{(1-\xi)_+} - 
\frac{2(1+\xi^2)\ln \xi}{(1-\xi)} - \frac{2(1+\xi^2)\ln \lambda }{(1-\xi)_+}\nonumber\\
&\quad-2 (1+\xi^2)\[ \frac{\ln (1-\xi)}{1-\xi} \]_+ - \delta(1-\xi) [3 \ln \lambda + 5 
+ 2 \pi^2/3]   \Big\} \, .
\eea
Note that, as pointed out in Ref.~\cite{Kretzer:1998ju}, the previous
expressions presented in Ref.~\cite{Hoffmann:1983ah} are affected by a
typo, and also differ
by terms which vanish in the
massless limit.

\bibliography{icfin}

\end{document}